# From Classical Machine Learning to Emerging Foundation Models: Review on Multimodal Data Integration for Cancer Research

**Amgad Muneer [1], Muhammad Waqas [1], Maliazurina B Saad [1], Eman Showkatian [1], Rukhmini Bandyopadhyay[1], Hui Xu[1], Wentao Li[1], Joe Y. Chang [3], Zhongxing Liao [3], Cara Haymaker [4], Luisa Solis Soto [4], Carol C Wu [5], Natalie I Vokes [2], Xiuning Le[2], Lauren A Byers [2], Don L Gibbons [2], John V Heymach [2], Jianjun Zhang [2], Jia Wu [1,2,*]**

[1] Department of Imaging Physics, The University of Texas MD Anderson Cancer Center, Houston, TX 77030, USA.
[2] Department of Thoracic/Head and Neck Medical Oncology, The University of Texas MD Anderson Cancer Center, Houston, TX 77030, USA.
[3] Department of Thoracic Radiation Oncology, The University of Texas MD Anderson Cancer Center, Houston, TX 77030, USA.
[4] Department of Translational Molecular Pathology, The University of Texas MD Anderson Cancer Center, Houston, TX 77030, USA.
[5] Department of Thoracic Imaging, The University of Texas MD Anderson Cancer Center, Houston, TX 77030, USA.

* Correspondence: jwu11@mdanderson.org

**Abstract:** Cancer research is increasingly driven by the integration of diverse data modalities, spanning from genomics and proteomics to imaging and clinical factors. However, extracting actionable insights from these vast and heterogeneous datasets remains a key challenge. The rise of foundation models (FMs) large deep-learning models pretrained on extensive amounts of data serving as a backbone for a wide range of downstream tasks—offers new avenues for discovering biomarkers, improving diagnosis, and personalizing treatment. This paper presents a comprehensive review of widely adopted integration strategies of multimodal data to assist advance the computational approaches for data-driven discoveries in oncology. We examine emerging trends in machine learning (ML) and deep learning (DL), including methodological frameworks, validation protocols, and open-source resources targeting cancer subtype classification, biomarker discovery, treatment guidance, and outcome prediction. This study also comprehensively covers the shift from traditional ML to FMs for multimodal integration. We present a holistic view of recent FMs advancements and challenges faced during the integration of multi-omics with advanced imaging data. We identify state-of-the-art FMs, publicly available multimodal repositories, and advanced tools and methods for data integration. We argue that current state-of-the-art integration methods provide the essential groundwork for developing the next generation of large-scale, pre-trained models poised to further revolutionize oncology. To the best of our knowledge, this is the first review to systematically map the transition from conventional ML to advanced FM for multimodal data integration in oncology, while also framing these developments as foundational for the forthcoming era of large-scale AI models in cancer research. The GitHub repo of this project available at https://github.com/WuLabMDA/Medical-Foundation-Models.

## 1. Introduction

Cancer is a leading cause of morbidity and mortality worldwide [1], characterized by its complexity, heterogeneity, and adaptability [2]. Its progression is governed by intricate interactions among genetic, epigenetic, proteomic, and metabolic networks, making it one of the most challenging diseases to diagnose, prognosticate, and treat effectively [3, 4]. In the era of precision medicine, there is an urgent need for integrative approaches that can unravel these complexities and provide actionable insights into individual tumors' unique biological and phenotypic characteristics [5, 6].

Recent advancements in high-throughput technologies have ushered in the age of multi-omics [7], encompassing genomics [8], transcriptomics [9], proteomics [10], metabolomics [11], and epigenomics [12]. These technologies generate massive datasets that hold the key to understanding cancer at a molecular level, enabling researchers to identify biomarkers [13], elucidate disease mechanisms [14], and predict therapy responses [15]. Similarly, imaging modalities [16] have become indispensable tools in cancer diagnostics [17-19] and treatment planning [20, 21]. These modalities provide spatial and temporal information about tumor morphology and the surrounding microenvironment [22], supplementing the molecular insights derived from omics data [7-12]. Building on this, quantitative imaging has emerged as a systematic framework for extracting high-dimensional features from clinical images. Radiomics derives engineered or deep features from routine computed tomography (CT), magnetic resonance imaging (MRI), and positron emission tomography (PET) scans, capturing tumor shape, intensity, and texture [23]. Pathomics extends this paradigm to whole-slide histopathology, using patch-level descriptors or deep embeddings to quantify tissue architecture, cellular morphology, and the spatial organization of the tumor microenvironment [24].

However, the integration and analysis of multimodal data present formidable challenges due to the complexity and diversity of the datasets involved [25] (**Figure 1**). Molecular data are structured yet high-dimensional and complex [26], whereas imaging data (radiology and histopathology) are high-resolution, spatially rich representations that require specialized feature extraction pipelines to become amenable to statistical modeling [27]. In this review, we use "integration" in a broad but precise sense to include (i) vertical integration, where multiple data types (e.g., genomics, transcriptomics, radiomics, pathomics, clinical variables) are linked at the level of the same patient or tumor; (ii) horizontal integration, where cohorts measured with overlapping but non-identical modalities are combined; and (iii) cross-modal alignment of datasets that are only partially overlapping in samples or features, typically via a shared latent representation or alignment objective, in line with recent multi-omics taxonomies. Combining these datasets into unified analytical frameworks requires advanced computational tools capable of capturing intricate relationships across modalities, while handling differing dimensionalities, noise structures, and sampling schemes. Traditional statistical and computational approaches often fall short in handling this complexity, necessitating innovative solutions that can bridge the gap between diverse data types [28, 29].

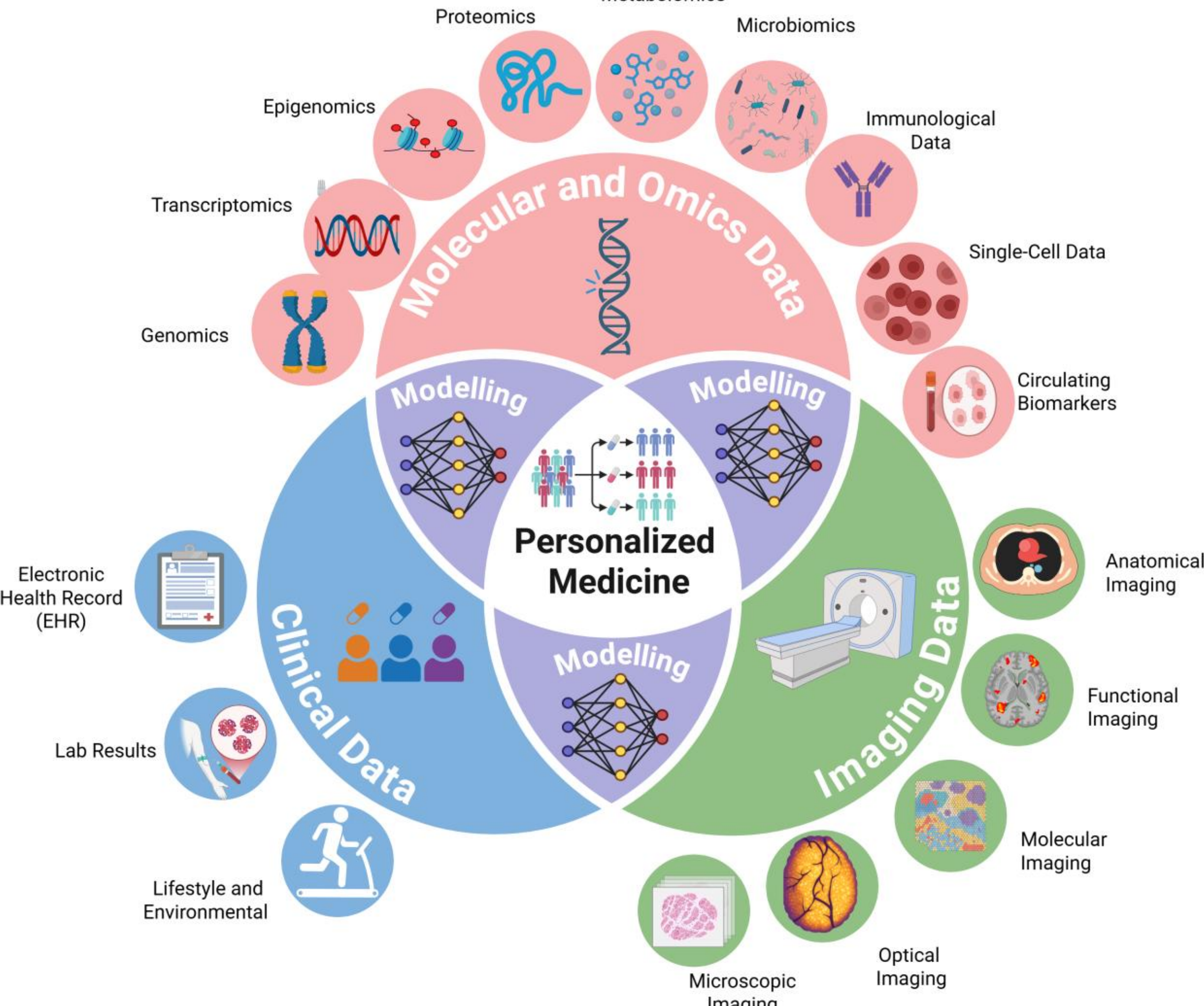

**Figure 1.** Multi-modal data integration towards personalized medicine.

Machine learning (ML), particularly its DL subfield, has revolutionized the integration and analysis of high-dimensional, multimodal cancer datasets [30]. ML algorithms, including support vector machines (SVM) and tree-based ensemble methods such as random forests (RF) and gradient-boosted decision trees (e.g., XGBoost), are highly effective in identifying patterns and making predictions from multi-omics data [31-33]. DL has further extended these capabilities, providing powerful tools to learn hierarchical representations directly from raw or minimally processed data and to capture complex nonlinear relationships; however, this flexibility also makes such models susceptible to overfitting and dataset-specific biases if they are not carefully regularized and rigorously validated [34, 35]. Advanced architectures like convolutional neural networks (CNNs) excel in extracting spatial features from imaging data [36], recurrent neural networks (RNNs) specialize in sequential data like temporal gene expression [37, 38], and transformers, with their ability to handle large-scale data and contextual dependencies, achieve outstanding results in cancer diagnosis, prognosis, and therapy response prediction [39, 40]. These AI-driven approaches are bridging the gap between molecular and phenotypic data, offering a deeper understanding of cancer biology.

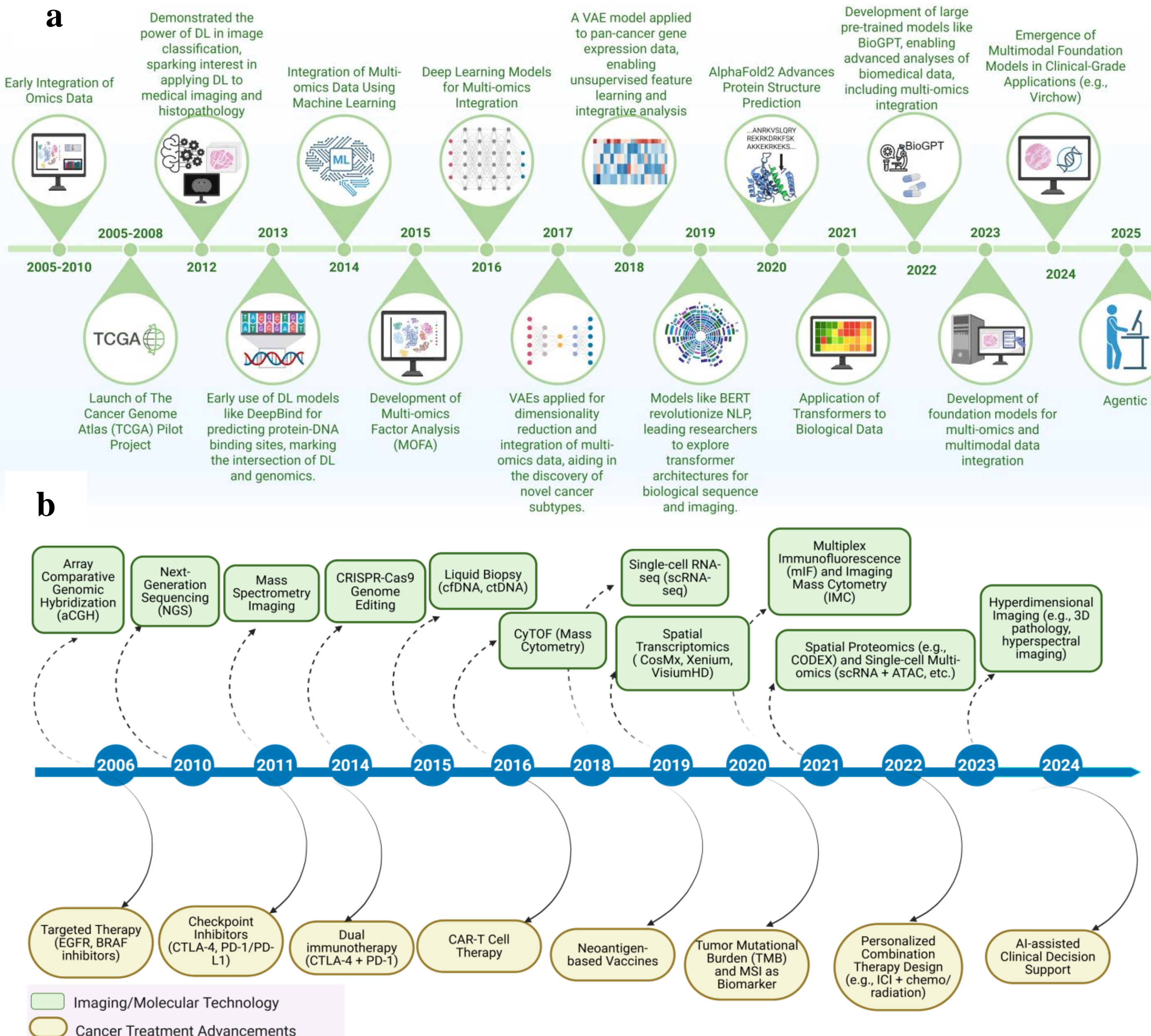

**Figure 2.** Timeline of (a) DL Advancements in Cancer Research through Multimodal Approaches; (b) Imaging/ molecular technology and cancer treatment.

Recently, the emergence of FMs has changed the landscape of AI. These models are trained on massive datasets, and sophisticated architectures are designed to generalize across a wide range of domain-specific tasks [41]. In oncology, they demonstrate remarkable capabilities in integrating and interpreting multimodal data, including omics profiles, radiological images, histopathological slides, and clinical records, revealing complex associations between molecular and phenotypic features [42-46]. The evolution of computational oncology is driven by a powerful synergy between advancements in AI (**Figure 2a**) and breakthroughs in molecular technology and cancer therapy (**Figure 2b**). The data explosion from technologies like Next-Generation Sequencing (NGS) and spatial omics has fueled the development of sophisticated DL models. In turn, these AI advancements now enable personalized therapies and AI-assisted clinical decision support, creating a powerful feedback loop of innovation

Although previous reviews have examined specific aspects of multimodal integration, most have focused narrowly on particular methodologies, individual cancer types, or single data modalities. As a result, an important gap remains: a comprehensive synthesis that evaluates state-of-the-art ML techniques, emerging foundation models, and their integrated application across diverse cancer contexts. Given the rapid evolution and growing complexity of these technologies, there is an urgent need for a holistic review that not only summarizes recent advances but also clarifies their translational implications and delineates unresolved challenges. In this context, our review makes three main contributions. First, we provide a systematic, cancer-focused overview of multimodal integration methods spanning the full spectrum from classical ML through DL to contemporary FMs, explicitly mapping how model design, data requirements, and validation strategies have evolved over time. Second, we connect methodological innovation to practice by jointly synthesizing algorithmic developments and multimodal integration strategies in relation to concrete clinical endpoints (diagnosis, prognosis, treatment response, and workflow optimization) thereby highlighting where these techniques already influence, or are poised to influence, patient care. Third, informed by close collaboration with clinicians, we articulate a structured set of open challenges and future directions, encompassing data-centric issues, model-centric limitations, and clinical implementation barriers. To this end, these elements position this work as a rigorous, cancer-focused critical appraisal of multimodal integration methods from classical ML to modern FMs, designed to support ongoing research efforts and to help steer the next generation of computational oncology.

## 2. Methods

### *2.1. Evidence Acquisition and Search Strategy*

To assemble the evidence base for this review, we performed a structured literature search and screening process focused on multimodal cancer studies that use ML, DL, or FMs. This section details the search strategy, eligibility criteria, and selection workflow for the meta-analysis in section 4.1 using PRISMA methodology (**Figure 3**). In parallel, we conducted a complementary narrative scoping of recent cancer-relevant FMs including preprints to construct the FM taxonomy in Section 5.4, as many of these models do not yet satisfy the strict inclusion criteria applied to multimodal cancer studies.

A comprehensive literature search was conducted in the National Library of Medicine's PubMed database (https://pubmed.ncbi.nlm.nih.gov). We considered articles published between 1 January 2019 and 12 November 2024, to capture contemporary ML/DL and FM–based multimodal methods. The core search string combined terms for multimodal/multi-omics data, cancer, and AI methods: (multiomics OR "multi-omics" OR "multi omics" OR "integrative omics" OR "integrated omics" OR multimodal) AND (cancer OR tumor OR tumour) AND ("machine learning" OR "deep learning" OR "artificial intelligence" OR "foundation model" OR "neural network") Reference lists of recent reviews on multi-modal integration and medical FMs were also screened to identify additional eligible studies not captured by the database query. The overall identification and screening process is summarized in **Figure 3**.

### *2.2. Eligibility Criteria and Screening*

We applied predefined inclusion and exclusion criteria to focus on original research articles that use ML/DL or FMs for multimodal cancer analysis. The following criteria were applied:

Inclusion Criteria

- Original research articles (not reviews, editorials, or perspectives).
- Human or clinically oriented cancer studies that integrate two or more data modalities (e.g., genomics, transcriptomics, epigenomics, proteomics, histopathology, radomics, clinical variables).

- Use of ML, DL, or FM methods as a central component of the analysis (e.g., for diagnosis, prognosis, risk stratification, response prediction, or treatment planning).
- Full-text available in English.

Exclusion Criteria

- Studies on multi-omics or multimodal integration outside of cancer domain.
- Studies focusing on multi-omics/multimodal data but were not in cancer research were not considered.
- Non-peer-reviewed or low-quality sources (e.g., theses, non-archival workshop abstracts) when they did not provide sufficient methodological detail or validation.
- Non-English publications and duplicate or superseded versions of the same study.

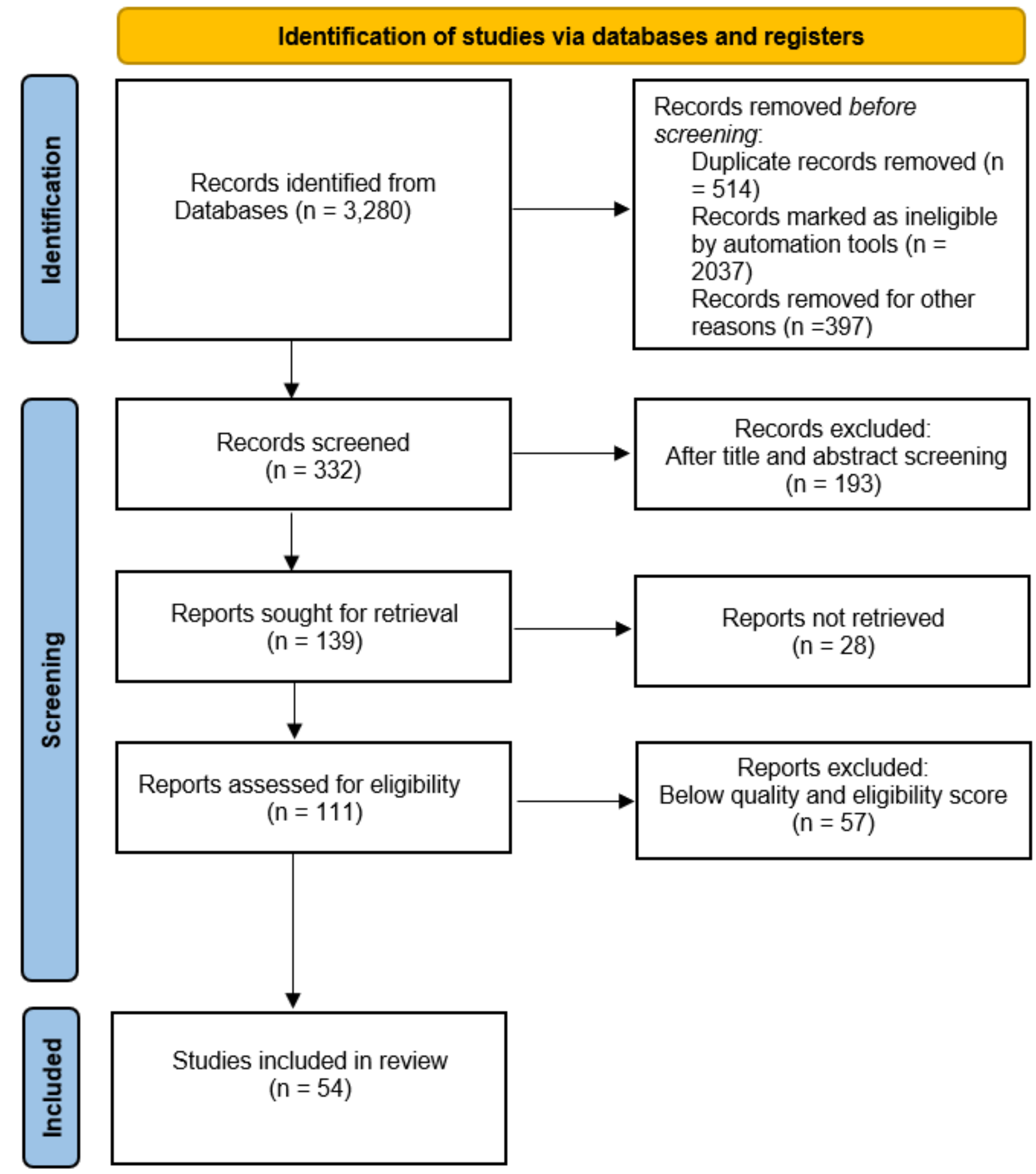

**Figure 3:** PRISMA methodology for study selection.

Using PubMed and the search strategy described above, we initially identified 3,280 records. Before screening, 514 duplicates, 2,037 records flagged as ineligible by PubMed's automated tools, and 397 records removed for other reasons were excluded, leaving 332 unique records for title–abstract screening. Of these, 193 were excluded as not meeting the multimodal cancer and ML/FMs focus, and 139 full texts were sought for retrieval. Twenty-eight reports could not be retrieved (e.g., inaccessible for full text or incomplete records), leaving 111 articles for full-text assessment. At this stage, 57 articles were excluded for "below quality and eligibility score" because they lacked essential details on data characteristics, integration strategy, or model validation, or performed only minimal integration without a clear multimodal component. The remaining 54 studies constitute the core evidence base for the meta-analysis discussed in Table 4.

*2.3. Curation of FMs Taxonomy*

Because many cancer-relevant FMs are reported as preprints (e.g., arXiv or BiorXiv) and are not yet officially published, they did not satisfy all the above eligibility criteria. To capture this rapidly evolving space, we therefore performed a complementary narrative scoping review for Section 5.4. We manually searched PubMed, google scholar, arXiv, and BiorXiv venues using combinations of ("foundation model*" OR "foundation models" OR "pre-trained model*" OR "pretrained model*") AND (omics OR "multi-omics" OR "multi omics" OR multiomics OR "single-cell" OR "single cell") AND (multimodal* OR "multi-modal" OR "cross-modal") AND ("digital pathology" OR "computational pathology" OR "whole slide" OR "whole-slide" OR WSI OR histopatholog* OR CT OR "computed tomography") AND (cancer OR tumour* OR tumor OR oncology). Candidate models were included in the taxonomy if they: (i) provided a general-purpose pretrained representation (omics, pathology, radiology, or multimodal) relevant to cancer; and (ii) were accompanied by at least one downstream task or application in oncology or have been employed in cancer domains.

**3. Multi-modal Integration Methods**

The landscape of multimodal integration methods can be systematically understood through a two-tiered classification framework that separates foundational architectural choices from advanced algorithmic strategies. The foundational tier is defined by the timing of integration: early fusion merges raw or processed features at the input level; intermediate fusion combines learned representations within the model architecture; and late fusion ensembles the outputs or decisions from models trained on individual data modalities. Distinct from these architectural blueprints are the advanced, algorithm-based methods that define how integration is computationally performed. This diverse and rapidly evolving category includes hierarchical [47], attention-based [48], multi-view learning [49], graph-based [50], and correlation-based strategies [51], which are increasingly employed to move beyond simple feature concatenation and model the profound complexity of biological systems [52, 53].

Although many contemporary architectures, particularly large multimodal FMs [54-57], combine several of these mechanisms within a single network, the two axes we emphasize: (i) the timing of fusion (early/intermediate/late) and (ii) the dominant computational strategy (e.g., attention, graphs, multi-view learning, correlation-based methods) remain useful organizing principles. Rather than prescribing rigid categories, this framework is intended as an extensible lens through which future models, including increasingly unified FMs, can be decomposed and compared. In this review, each modality can correspond to a molecular omics layer (e.g., genomics, transcriptomics, proteomics), an imaging-derived representation (radiomics, pathomics), or clinical variables, all of which must be coherently integrated within a common computational framework. Historically, many of these integration strategies first appeared in classical ML pipelines (e.g., early fusion with SVMs or random forests, late-fusion ensembles of modality-specific classifiers) and were later absorbed into DL architectures and, more recently, FMs-based systems, a trajectory that we analyze in greater depth in Section 5.

A second design axis concerns whether data are matched or unmatched. In matched designs, all modalities are measured on the same patient, sample, or cell, enabling direct vertical integration across omics and imaging layers. In unmatched settings, different cohorts, time points, or institutions contribute to non-overlapping subsets of modalities (e.g., one cohort with RNA-seq only and another with histopathology only). Here, integration must rely on alignment via shared features (genes, pathways, cell types), statistical matching, or latent-space coupling rather than straightforward sample-wise concatenation. This distinction is critical: many classical early-fusion pipelines implicitly assume fully matched data, whereas modern deep and foundation-model approaches increasingly target partially matched or entirely unmatched datasets.

Orthogonal to structure is modelling intent. Multimodal models may be trained as single-task predictors (e.g., survival in one cancer type), as multi-task systems optimizing several related endpoints, or as general-purpose pre-trained backbones later adapted via fine-tuning or prompting. The same fusion pattern, for example, intermediate, representation-level fusion with attention can therefore underpin narrowly focused classifiers [58] or broad FMs used as shared encoders. Throughout this section we focus on method classes; modelling intent is revisited explicitly in the FM discussion (Section 5.4).

Finally, multimodal integration almost always operates on feature representations rather than raw data streams. For molecular omics, these representations include: (i) handcrafted descriptors (e.g., MACCS/ECFP fingerprints, curated gene signatures, pathway scores); (ii) model-derived or knowledge-enriched features (e.g., 3D structural embeddings from AlphaFold, network-derived centrality or module scores, outputs of prior prognostic models); and (iii) unsupervised or self-supervised embeddings, such as autoencoder bottlenecks, contrastive-learning representations, or latent spaces from FMs. Imaging modalities follow a parallel progression from handcrafted radiomics features texture [23] to CNN or transformer embeddings and joint vision–language representations. The integration strategies surveyed below are largely agnostic to the specific feature type but differ in how effectively they can combine, regularize, and interpret these heterogeneous feature spaces.

### *3.1. Stage-Wise Fusion Methods*

Foundational integration strategies are categorized by the time point at which multimodal data are computationally fused. The simplest approach, early fusion, concatenates all feature vectors at the input level before they are fed into a unified model. Conversely, late fusion operates at the output level, where predictions from distinct, modality-specific models are combined for a final decision (**Figure 4**). Intermediate fusion offers a compromise, merging learned feature representations at an intermediate layer within a deep learning architecture. This allows initial layers to learn modality-specific features while subsequent layers perform joint analysis. Each strategy offers a different balance between model complexity, modularity, and the ability to capture cross-modal dependencies.

#### 3.1.1 Early (Feature-Level Integration)

Early fusion, or feature-level integration, concatenates features from all modalities into a single representation before learning begins [59]. If $X^{(1)}, X^{(2)}, \dots, X^{(m)}$ denote the extracted feature matrices from each modality, then the fused input is simply:

$$X_{\text{concat}} = [\, X^{(1)} \mid X^{(2)} \mid \cdots \mid X^{(m)} \,], \tag{1}$$

This integrated feature matrix is then passed to a model to predict the output of interest. Early fusion is conceptually simple and can capture fine-grained cross-modal interactions, but it suffers from two practical challenges: (i) very high dimensionality, which exacerbates overfitting, and (ii) the need for complete data, making the approach brittle when modalities are missing or only partially available [35].

#### 3.1.2. Intermediate (Representation-Level Integration)

Intermediate fusion first transforms each modality into a compact latent space before combining them. For each modality, an encoder $g^{(i)}(\cdot)$produces an embedding:

$$H^{(i)} = g^{(i)}(X^{(i)}), \tag{2}$$

These modality-specific embeddings are then merged typically by concatenation, averaging, or an additional fusion module, and passed to a predictive model. Intermediate fusion maintains modality-specific learning while enabling joint representation analysis. However, interpretability can suffer because latent spaces are not always biologically meaningful.

#### 3.1.3. Late (Decision-Level Integration)

Late integration, or decision-level integration, entails training separate models for each modality combines their predictions rather than their features [59, 60]. For modality i:

$$\hat{Y}^{(i)} = f_{\theta}^{(i)}\big(X^{(i)}\big), \tag{3}$$

The final prediction $\hat{y}$ is obtained by aggregating the individual predictions through an aggregation function $\varphi(:)$

$$\hat{Y} = \varphi\big([\hat{Y}^{(1)}, \hat{Y}^{(2)}, \dots, \hat{Y}^{(i)}]\big), \tag{4}$$

Late fusion is robust to missing modalities and modular across cohorts but cannot capture feature-level synergies because cross-modal interactions are only considered at the decision level.

Unfortunately, these predefined data fusions are often too rigid, as early fusion requires perfectly aligned data that is sensitive to missing modalities, while late fusion fails to uncover complex, synergistic interactions between the data sources. This static approach often forces a compromise, either losing granular details early on or missing crucial cross-modal correlations by waiting until the very end to integrate insights.

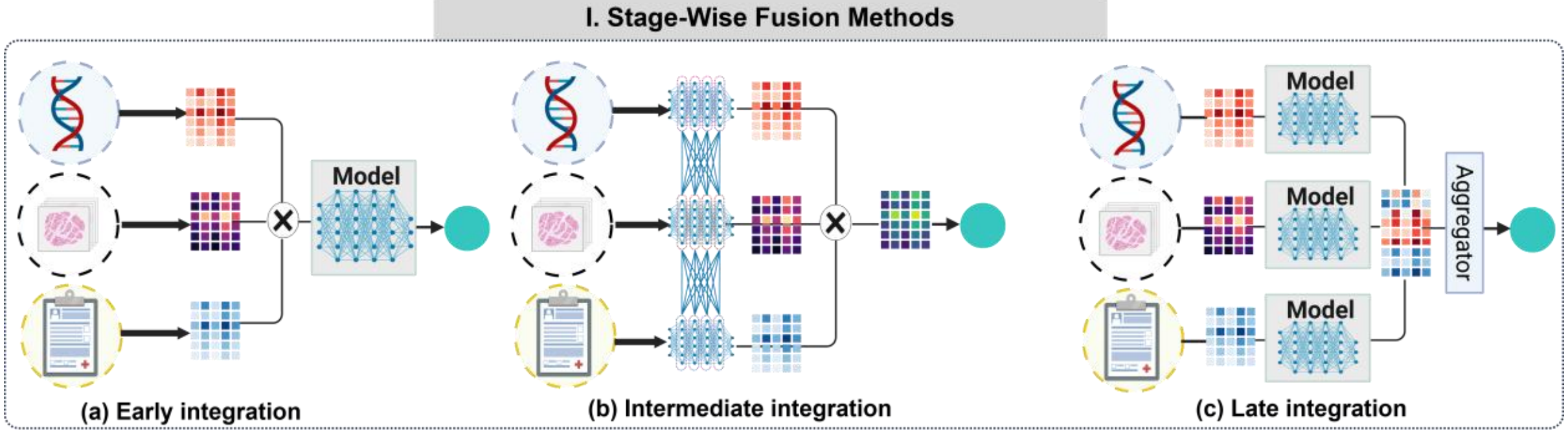

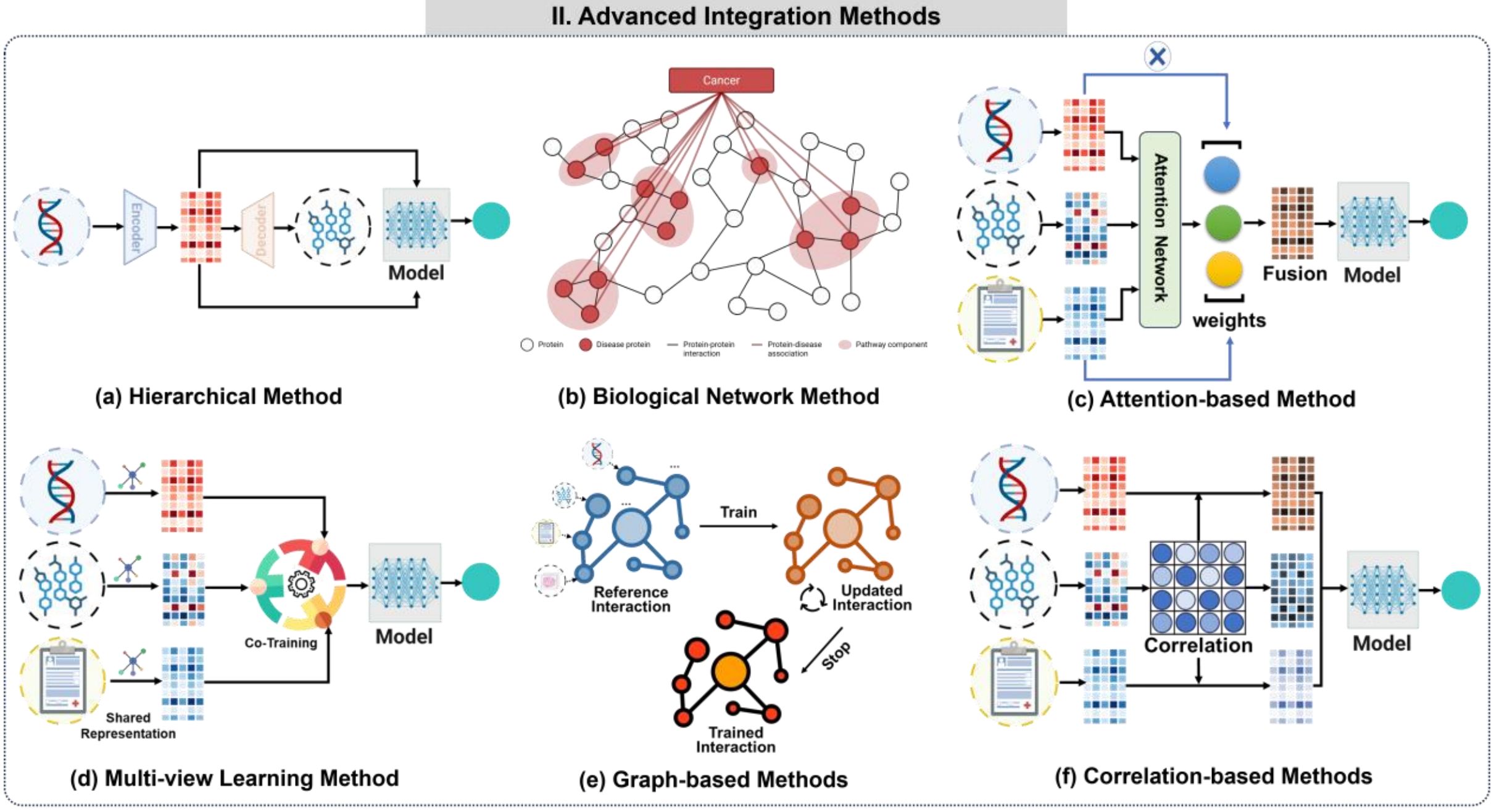

**Figure 4.** Methods of multi-modal data integration: (**a**) Stage-wise fusion; (**b**) advanced integration methods.

### *3.2. Advanced Integration Methods*

Beyond simple fusion, advanced integration methods employ distinct computational philosophies to harness multimodal data. To reflect known biological hierarchies and dependencies, hierarchical methods construct structured, multi-layered models. For dynamically identifying the most salient features within high-dimensional data, attention-based methods offer a powerful solution, assigning importance weights that also aid interpretability. Multi-view learning addresses the challenge of heterogeneous data by treating each modality as a separate "view," seeking to learn a shared representation that captures both consensus and complementary patterns. Graph-based methods, including graph neural networks (GNNs), are purpose-built to model the explicit relational structures and interactions inherent in biological networks. Lastly, to reduce dimensionality and discover co-variation patterns, correlation-based methods leverage matrix and tensor factorization or canonical correlation analysis to find shared latent variables.

#### *3.2.1 Hierarchical Methods*

Hierarchical integration methods leverage the established central dogma of molecular biology as a biologically informed blueprint for multimodal fusion. These approaches impose a directed flow of information across omics layers (genomics, transcriptomics, proteomics, and metabolomics) mirroring cellular regulatory cascades. Each omics dataset is first encoded into a latent representation following the hierarchical order of biological regulation. For the first layer, the latent representation is obtained using a feed-forward neural encoder:

$$H^{(1)} = f_1\left(X^{(1)}\right) \tag{5}$$

For each downstream biological level $\ell = 2, \dots, L$, the representation is conditioned on both its own features and the upstream encoded layer:

$$H^{(\ell)} = f_\ell\left(X^{(\ell)}, H^{(\ell-1)}\right) \tag{6}$$

To combine information across biological layers, hierarchical importance weights $\alpha^{(\ell)}$are computed using a neural scoring function $g_{\text{hier}}(\cdot)$applied to each layer's latent representation:

$$\alpha^{(\ell)} = \frac{\exp\left(g_{\text{hier}}\left(H^{(\ell)}\right)\right)}{\sum_{k=1}^{L} \exp\left(g_{\text{hier}}\left(H^{(k)}\right)\right)}, \text{s.t. } \alpha^{(\ell)} \geq 0,\ \sum_{\ell=1}^{L} \alpha^{(\ell)} = 1 \tag{7}$$

The hierarchical fusion representation is then obtained by weighing the latent representations according to their biological importance:

$$H_{\text{hier}} = \sum_{\ell=1}^{L} \alpha^{(\ell)} H^{(\ell)} \tag{8}$$

Models such as TranPo [47] exemplify this strategy by transferring gene-level information to protein-level representations, while Bayesian frameworks like iBAG [61] explicitly model regulatory relationships (e.g., copy-number variation → gene expression) within the hierarchical structure. Although biologically intuitive, hierarchical integration methods depend heavily on prior knowledge from pathway or interaction databases, which may limit performance when external annotations are incomplete or inconsistent.

*3.2.2 Biological Network-Based Methods*

Network-based integration methods project multi-modal data onto known biological interaction networks such as protein-protein interactions (PPIs)[62], gene regulatory networks (GRNs)[53], or metabolic pathways [52] enabling a holistic understanding of complex biological systems. This allows for the tracing of molecular perturbations across biological layers and the identification of influential network modules for biomarker discovery [63, 64]. The core limitation of this approach is its dependence on a 'ground truth' network; the resulting insights are therefore only as reliable as the underlying, often incomplete and biased, interaction map. Coupled with the inherent challenge of interpreting complex network topologies, ensuring the biological relevance of purely network-driven findings remains a key concern [29, 65]. Each omics dataset $X^{(i)}$is first mapped to an initial node feature representation using an encoder:

$$H^{(i)} = f_i(X^{(i)}) \tag{9}$$

To integrate regulatory and interaction structure, node features are propagated across the biological network. $G = (V, E)$using a graph-based message-passing rule:

$$H'_v = \sigma\Big(\sum_{u \in \mathcal{N}(v)} W\, H_u + b\Big), \forall v \in V \tag{10}$$

where $\mathcal{N}(v)$denotes the neighbors of node $v$, and $\sigma(\cdot)$is a nonlinear activation function.
A global network-level representation is then obtained by aggregating node features across the graph:

$$H_{\text{network}} = \text{POOL}(\{H'_v : v \in V\}) \tag{11}$$

Common pooling operators include mean pooling, sum pooling, or attention-based pooling. The final integrated representation $H_{\text{network}}$can then be used for prediction or downstream analysis.

*3.2.3 Attention-based Methods*

Attention mechanisms enhance neural networks by allowing the model to focus on the most relevant parts of the input data. In attention-based multi-modal integration, each omics dataset is projected to a fixed dimension latent space using a feed forward neural network $\boldsymbol{H}^{(i)} = f_i(X^i) \in \mathbb{R}^{n \times d_h}$. Given the latent representation of each omics data, an attention weights $\alpha^{(i)}$ are computed using neural network $g_{att}(\cdot)$ as:

$$\alpha^{(i)} = \frac{\exp\left(g_{att}\left(\boldsymbol{H}^{(i)}\right)\right)}{\sum_{j=1}^{m} \exp\left(g_{att}(\mathbf{H}^{(j)})\right)}, \text{ s.t. } \alpha_i \geq 0, \forall i \in \{1,2,\ldots,m\}, \text{ and } \sum_{i=1}^{m} \alpha_i = 1 \tag{12}$$

The attention weights determine the contribution of each omics modality to the final representation $H_{\text{attention}}$, which is a weighted sum of the individual representations:

$$H_{\text{attention}} = \textstyle\sum_{i=1}^{i} \alpha^{(i)} \boldsymbol{X}^{(i)}, \tag{13}$$

The attention-based methods offer adaptive integration while providing improved interpretability of how different omics data contributes to decision making process [66].

*3.2.4 Multi-view Learning Methods*

Multi-view learning provides a powerful conceptual framework for data integration, treating each omics modality as a distinct 'view' or perspective of the same underlying biological system [67, 68]. The fundamental characteristic of this integration approach is co-learning [49], where the models are trained on separate views, and predictions are combined to maximize the consensus among the models in semi-supervised settings. The reason for its success is discussed in [69, 70]. Multi-view learning is also extended with the kernel learning approach to integrate features from different views. DL-based approaches are also introduced to multi-view integration [34, 71]. Recent studies have focused on semi-supervised learning to capture both global structure and local dependencies across omics data layers [72-75]. Furthermore, graphs, autoencoder, and DL-based hybrid approaches also promise to capture local and global dependencies [76-80]. Despite their sophistication, these methods are often less effective when significant disagreement exists between views or when one modality is substantially more informative than others, as the drive consensus can suppress critical, unique signals. Furthermore, ensuring that the learned shared space is biologically meaningful, rather than a statistical artifact, remains a significant challenge. Each omics view $X^{(i)}$is first encoded into a latent representation using a view-specific encoder:

$$H^{(i)} = f_i(X^{(i)}) \tag{14}$$

A shared latent representation $Z$is then learned by minimizing disagreement across all views:

$$Z = \arg\min_{Z} \sum_{i=1}^{m} L(H^{(i)}, Z) \tag{15}$$

where $L(\cdot,\cdot)$is a divergence or consistency loss that measures alignment between a view and the shared space. The final integrated representation is obtained by fusing the aligned view-specific embeddings with learned view weights $\alpha^{(i)}$:

$$H_{\text{multi-view}} = \sum_{i=1}^{m} \alpha^{(i)} H^{(i)} \text{,s.t. } \alpha^{(i)} \geq 0,\ \sum_{i=1}^{m} \alpha^{(i)} = 1 \tag{16}$$

This shared embedding $H_{\text{multi-view}}$is then used for downstream prediction or clustering tasks.

*3.2.5 Graph-based Methods.*

Graph-based methods integrate the topological structure and inherent relationship of omics data by representing multi-modal data $\boldsymbol{X} \in \mathbb{R}^{n \times D}$ as a unified graph $G = (\boldsymbol{V}, \boldsymbol{E})$, where $\boldsymbol{V}$ correspond to biological entities such as genes or proteins, and edges $\boldsymbol{E}$ represent interactions or associations between them (*e.g.*, gene-gene interactions, regulatory links, protein-protein interactions). The connectivity of the graph is encoded by an adjacency matrix $\boldsymbol{A}$. Graph Neural Net GNNs then employ a layered architecture to learn node embeddings $\boldsymbol{H}^{(\boldsymbol{L})} = \{h_1^{(L)}, h_2^{(L)}, \ldots, h_n^{(L)} \in \mathbb{R}^{n \times d'}$, where the dimension of learned embeddings is denoted by $d'$and $L$ denotes the layer of GNN. For each GNN layer $l$ the embeddings $H^{(l+1)}$ are computed using message passing architecture:

$$\boldsymbol{H}^{(0)} = X,\ \boldsymbol{h}_i^{(0)} = \boldsymbol{x}_i, \tag{17}$$

$$h_i^{(l+1)} = \sigma\left(\boldsymbol{W}^{(l)} h_i^{(l)} + \textstyle\sum_{j \in \mathcal{N}(i)} \phi^{(l)}\left(h_i^{(l)}, h_j^{(l)}, e_{ij}\right)\right), \tag{18}$$

Where $\phi^{(l)}$ is a differentiable function, e.g., weighted sum, or multi-layer perception (MLP), $\mathcal{N}(i)$ is the set of neighbors of a node of node $i$, $\boldsymbol{W}^{(k)}$ are learnable weight, and $\sigma$ is a non-linear activation function. After layer $L$, the final node embeddings $\boldsymbol{H}^L$ can be used to perform prediction.

$$\hat{Y} = f_\theta(\boldsymbol{H}^L), \tag{19}$$

Apart from single unified graph representations, recent advances have moved to more powerful heterogeneous information networks (HINs)[50]. This strategy first models each omics dataset as its own graph and then fuses them by defining different types of edges or meta-relations that explicitly annotate the specific biological interactions connecting nodes across these otherwise disparate graphs.

*3.2.6 Correlation-based Methods*

Correlation-based multi-modal data integration methods aim to capture correlations and quantify phenotypic traits, disease progression, and therapeutic responses. Early integration strategies, such as Partial Least Squares Regression (PLSR), focused on discovering latent variables by maximizing covariance between linear projections of omics data [81]. The foundational approach to correlate genomic variants with transcriptomic.[82], this method links single nucleotide polymorphisms (SNPs) to gene expression across the tissue. The correlated analysis of genomic and transcriptomic can reveal the critical role of tumor evolution and chemotherapy-induced mutagenesis in driving molecular complexity [51]. Similarly, the Cancer Genome Atlas (TCGA) [83] correlated multi-modal data to categorize molecular subtypes, including lung and breast cancer. GTEx Consortium [84] associates genetic variants with gene expression levels to provide critical insights into tissue-specific regulatory mechanisms**.** Canonical Correlation Analysis (CCA) is a widely adopted pivotal method for uncovering linear relationships across multi-modal datasets [85]. In this context, CCA learns projection vectors that maximize correlation between transformed omics datasets, thereby identifying shared patterns across biological layers and improving interpretability of cross-omics relationships. Given two omics datasets $X^{(i)}$and $X^{(j)}$, CCA seeks linear projections:

$$u = w^{(i)T} X^{(i)}, v = w^{(j)T} X^{(j)} \tag{20}$$

The objective of CCA is to find the projection vectors $w^{(i)}$and $w^{(j)}$that maximize the correlation between the projected variables $u$and $v$:

$$\max_{w^{(i)}, w^{(j)}} \operatorname{corr}(u, v) = \frac{\operatorname{cov}(u, v)}{\sqrt{\operatorname{var}(u) \operatorname{var}(v)}} \tag{21}$$

The resulting shared latent representation capturing cross-modal covariance is expressed as:

$$H_{\text{corr}} = [\, u,\, v \,] \tag{22}$$

This representation $H_{\text{corr}}$can then be used for prediction, clustering, or downstream integrative analyses.

Sparse and multiple extensions of canonical correlation analysis such as sparse CCA (SCCA), multiple CCA (MCCA), sparse group CCA (SGCCA), and its DIABLO variant (Data Integration Analysis for Biomarker discovery using Latent variable approaches for 'Omics studies) were developed to handle large-scale multi-omics by imposing sparsity, using covariance rather than correlation, and enabling feature selection across modalities [86-89]. To better capture non-linear relationships, deep CCA (DCCA) and its multi-view generalizations GCCA and DGCCA, including supervised variants, employ deep networks to learn shared latent representations across omics data [90-92]. More recently, advanced DL approaches have shifted multimodal integration toward learning latent cross-modal features directly: self-attention–based transformers uncover subtle dependencies across omics layers for tasks such as survival analysis in lung cancer [48], are combined with GNNs for disease classification [93], and are integrated with self-supervised objectives like contrastive learning for broad disease modeling [94]; in single-cell profiling, self-attention–driven generative models now set the state of the art for aligning modalities and annotating cell states [95].

### *3.3. Comparative Appraisal of Multimodal Integration Strategies and Recommended Use-Cases*

Across the nine multimodal integration methods discussed above, each offers distinct advantages, limitations, and assumptions that determine its suitability for different biological and clinical tasks (**Table 1**). Early fusion is simple and captures fine-grained cross-modal interactions but struggles with high dimensionality and missing data, whereas intermediate fusion mitigates these issues through representation learning yet sacrifices some interpretability. Late fusion [59, 60] excels when modalities are incomplete or heterogeneous but cannot model feature-level synergies . Hierarchical models leverage known regulatory order, providing mechanistic interpretability [47], but depend heavily on the correctness of biological priors. Graph-based approaches uniquely incorporate interaction networks and pathway structure, making them ideal when relational information is crucial [50], although their performance is constrained by the completeness and accuracy of curated biological graphs. Multi-view learning enforces consensus across modalities and is effective when views are complementary but may suppress unique modality-specific signals when disagreement exists [69, 70]. Correlation-based methods [51, 82] offer interpretable, statistically grounded insights into shared variation but are limited in capturing nonlinear relationships and require paired samples across modalities. Imaging–omics fusion integrates spatial information from radiology or pathology with molecular data, enabling spatial phenotyping and biomarker localization, but it requires large, high-quality imaging datasets and careful cross-modal alignment. Finally, multimodal FMs provide flexibility, capable of integrating images, omics, and text, but are computationally. Together, these diverse methods illustrate a spectrum of trade-offs between interpretability and flexibility, data requirements and robustness, reliance on biological priors and capacity for discovery, highlighting that method selection must be guided by data characteristics, biological assumptions, and the specific goals of the analysis.

**Table 1**. Comparative strengths, limitations, and recommended use-cases for multimodal integration strategies.

| Integration Method | Strengths | Limitations | Most Appropriate Use-Cases |
| --- | --- | --- | --- |
| Early Fusion (Feature-Level) | Simple; captures cross-modal interactions; minimal architectural complexity | High dimensionality; prone to overfitting; requires complete data; low interpretability | Large, aligned datasets; when cross-modal feature interactions are important |
| Intermediate Fusion (Representation-Level) | Reduces dimensionality; modality-specific representation learning; scalable | Loss of interpretability; dependent on learned representations; complex tuning | High-dimensional heterogeneous data requiring joint latent representations |
| Late Fusion (Decision-Level) | Robust to missing modalities; modular; more interpretable | Cannot model feature-level synergies; risk of modality dominance | Clinical settings with heterogeneous or incomplete multimodal data |
| Hierarchical Methods | Leverage biological priors; interpretable regulatory flow; mechanistically grounded | Dependent on accuracy of biological pathways; limited flexibility; error propagation | When regulatory directionality (DNA→RNA→protein) is known and meaningful |
| Graph-Based Integration | Captures network topology; pathway-aware; powerful relational modeling | Dependent on curated network quality; computationally intensive; complex interpretation | Pathway analysis, network biology, regulatory module discovery |
| Multi-View Learning | Uses complementary perspectives; encourages consensus; noise-robust | Fails with modality disagreement; may suppress unique signals; require sample matching | Complementary omics layers describing shared biology |
| Correlation-Based Methods (CCA, PLSR) | Interpretable; identifies shared variation; statistically grounded | Primarily linear; sensitive to noise; requires paired data | Linking omics layers; shared covariance discovery; exploratory analysis |
| Imaging–Omics Fusion (Spatial & Non-Spatial) | Captures spatial context; phenotype localization; attention improves interpretability | Requires large imaging datasets; alignment issues; computationally heavy | Radiopathomics, pathomics, spatial biomarker discovery |
| Multimodal FMs | Highly flexible; handle unpaired/incomplete data; powerful joint representations | High computational cost; reduced interpretability; requires large pretraining corpora | Large-scale multimodal diagnosis, prognosis, and biomarker discovery |

## 4. Tools and Applications for Multi-modal Data Integration

### *4.1. Multimodal Data Bank and Cancer Applications*

The integration of multi-modal data has become a cornerstone in advancing cancer research, mostly in the discovery of biomarkers that are prognostic and predictive of disease outcomes. Leveraging multiple omics layers (**Figure 5**) provides a holistic molecular understanding of cancer biology. Each omics type contributes unique insights such as genomics identifies genetic variations linked to disease and treatment response; proteomics explores protein expression dynamics and network configurations; transcriptomics captures RNA-level intermediaries of gene expression; metabolomics uncovers biochemical imbalances associated with tumorigenesis; and interact-omics maps functional protein-protein interactions critical to cellular processes. Emerging studies highlight the power of multi-omics in identifying robust biomarker discovery [81, 96, 97], cancer subtyping [98], therapeutic target identification, drug, and therapy response prediction [15, 47, 99], immune checkpoint, and PD-L1 prediction [100], recurrence and relapse prediction [101, 102], disease progression modeling [103], synthetic lethality detection. On the clinical side, the integrative methods identified in this review are already beginning to influence the translational pipeline by moving beyond purely methodological contributions to address concrete needs in diagnosis, prognosis, treatment selection, and workflow optimization [104].

To improve diagnostic accuracy and staging, integrative models are enabling earlier and more accurate cancer diagnosis and staging. For example, the fusion of deep features from whole-slide images (WSIs) and MRIs in [105] provides a non-invasive method for the early diagnosis of prostate bone metastasis with high Area Under the Curve (AUC: 0.85-0.93). In colorectal cancer, combining pathomics from slides with clinical markers significantly improves staging, with a combined model achieving an AUC of 0.814 on test data. Furthermore, ML analysis of cfDNA liquid biopsy data shows promises for non-invasive diagnosis of pediatric sarcomas with an AUC up to 0.97 [106]. Collectively, these studies illustrate how multimodal integration can improve early detection and staging, with direct implications for treatment planning and patient counseling.

For enhancing prognostication and risk stratification, a primary application of these models is the generation of more precise prognostic predictions, allowing for better patient risk stratification. Multi-omics models like Tumor Multi-Modal Pre-trained Network (TMO-Net) [107] and GNNs provide robust survival predictions that outperform single-data-type approaches. Multimodal fusion models are particularly impactful; Pathomic Fusion, which combines histology and genomics, demonstrated a concordance index (C-index) of up to 0.85 for predicting survival outcomes [108]. Similarly, integrating WSIs with clinical data using multimodal AI led to a hazard ratio of 2.33 for predicting distant metastasis in prostate cancer, directly informing patient prognosis. In principle, such models could support decisions about adjuvant therapy, intensity of surveillance, and enrollment in high-risk clinical trials, although most existing evidence remains retrospective [96].

Moreover, we found that some models are pivotal in advancing personalized medicine by predicting responses to specific therapies. A model integrating PET/CT imaging and clinical data can predict PD-L1 expression in Non-Small Cell Lung Cancer (NSCLC) with high accuracy (AUC: 0.82-0.89), which is critical for guiding immunotherapy decisions [100]. In breast cancer, a "radiopathomics" approach fusing MRI and WSI data successfully predicts chemotherapy response with an AUC between 0.81 and 0.86 [109]. Another model combining MRI, WSIs, and clinical data achieved a C-index of 0.860 for predicting biochemical recurrence after prostatectomy, helping to identify patients who may require adjuvant therapy [101]. These examples show that multimodal integration is not only improving risk prediction in abstract terms but also aligning predictions with concrete therapeutic decisions such as immunotherapy eligibility, chemotherapy benefit, and the need for intensified follow-up or additional treatment.

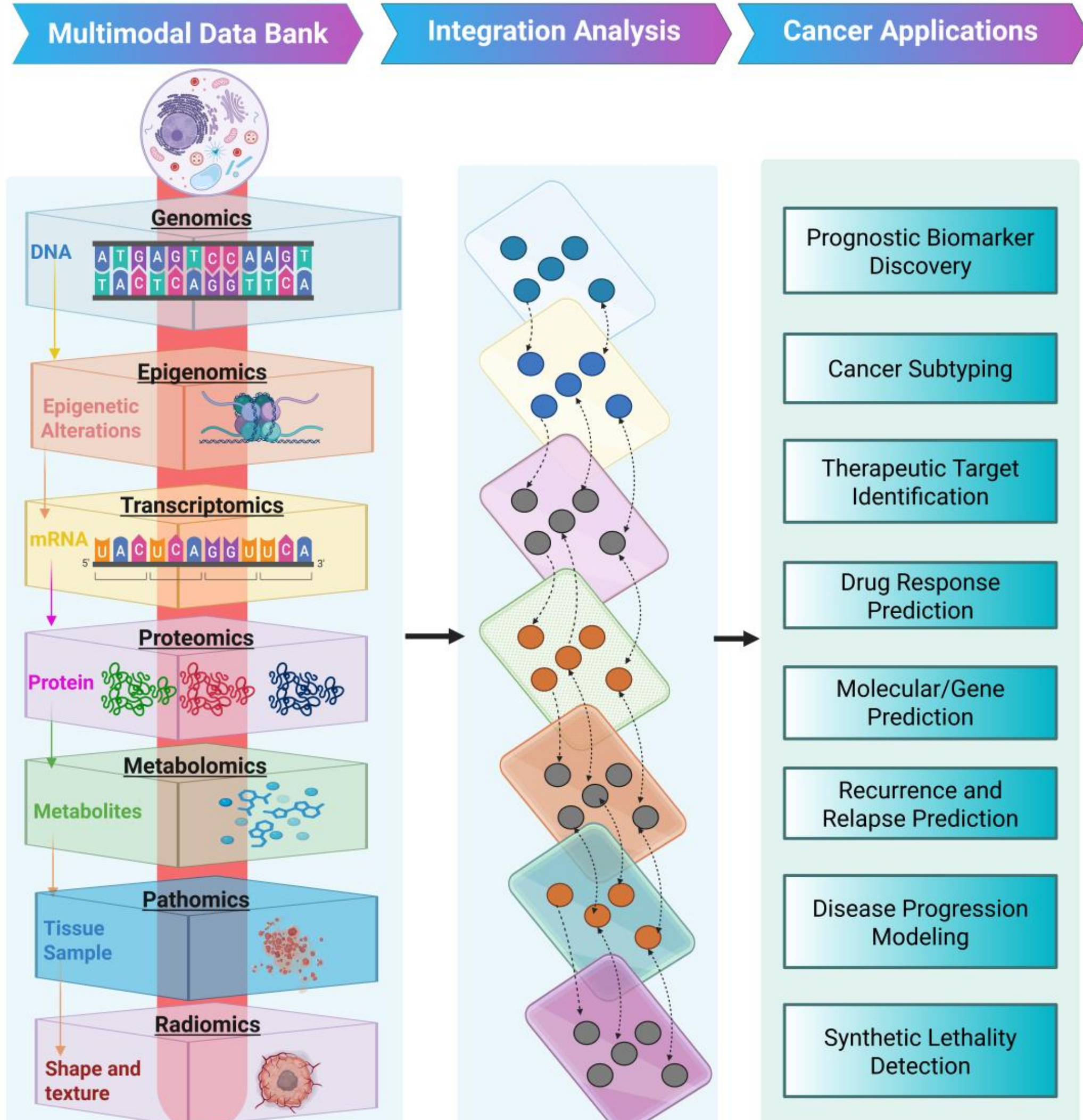

**Figure 5.** Multi-modal integration pipeline from modality data bank to cancer applications.

Additionally, AI-driven integration is also streamlining clinical processes. For instance, a 3D ResNeXt model that registers MRI and ultrasound images has been shown to reduce targeting error in prostate cancer biopsies by up to 62%, enhancing the precision of a routine clinical procedure [110]. Another model provides automated quality assessment for PET/CT scans, ensuring data integrity for clinical use with high inter-rater reliability (Kappa: 0.78-0.80) [111]. The fusion of endoscopy images and text reports via CNNs and Word2Vec has led to a highly accurate (95.3%) automated screening tool for upper gastrointestinal cancer, demonstrating the potential to improve efficiency in large-scale screening programs [112]. These workflow-oriented applications primarily impact healthcare efficiency and standardization, reducing manual workload, improving consistency across operators and centers, and potentially shortening time-to-diagnosis.

Several publicly available repositories provide comprehensive multimodal datasets (genomic, transcriptomic, proteomic, epigenomic, metabolomic, pathomics, radiomics and clinical metadata) across a wide range of diseases (**Table 2**). However, as discussed in Sections 5.2 and 5.3, realizing their full impact on patient care will require not only methodological innovation but also rigorous evaluation of clinical benefit, reproducibility, and equity.

*4.2. Conceptual Overview of Integration Approaches and Their Practical Characteristics*

To complement the integration framework described in Section 3, we summarize the major classes of multimodal fusion strategies using a conceptual, task-oriented perspective (**Table 3**). This table highlights how each integration category aligns with typical biomedical objectives, interpretability considerations, and input requirements. This structured overview provides readers with practical guidance on selecting an appropriate integration paradigm while maintaining generality across diverse data modalities.

**Table 2.** Comprehensive Multi-Modal Data Repositories

| DATA REPOSITORY | WEB LINK | DISEASE | Types Of Multi-Omics Data Available |
|---|---|---|---|
| The Cancer Genome Atlas (TCGA) | https://www.cancer.gov/tcga | Multiple cancer types (e.g., lung, breast, prostate, colorectal, etc.) | Genomics, Transcriptomics, Epigenomics, Proteomics, Clinical metadata. |
| Gene Expression Omnibus (GEO) | https://www.ncbi.nlm.nih.gov/geo/ | Various cancer types (breast, lung, colorectal, leukemia, prostate, melanoma, ovarian, etc.) | Genomics, Transcriptomics |
| Metabolomics Workbench | https://www.metabolomicsworkbench.org/ | Breast, lung, colorectal, prostate, hematologic malignancies, ovarian and other cancer types | Metabolomics (LC-MS, GC-MS, NMR) |
| Human Protein Atlas (HPA) | https://www.proteinatlas.org/ | Over 20 major cancers (breast, lung, colorectal, prostate, liver, ovarian, glioma, etc.) | Proteomics, transcriptomics (RNA-seq quantification), spatial proteomics, single-cell RNA-seq, clinical survival data |
| EGA (European Genome-Phenome Archive) | https://ega-archive.org/ | Breast, colorectal, prostate, lung, hematologic malignancies and other cancers. | Genomics, Epigenomics, Clinical metadata |
| ArrayExpress | https://www.ebi.ac.uk/arrayexpress/ | Breast, lung, colorectal, prostate, hematologic malignancies, ovarian, melanoma and other cancers | Transcriptomics, Epigenomics |
| cBioPortal | https://www.cbioportal.org/ | Multiple cancer types (e.g., breast, prostate, glioblastoma, leukemia) | Genomics, Transcriptomics, Clinical metadata. |
| ProteomicsDB | https://www.proteomicsdb.org/ | Various (e.g., cancer-focused proteomic and PTM profiles, drug–target interaction, disease biomarker and therapeutic target discovery) | Proteomics |
| ENCODE (Encyclopedia of DNA Elements) | https://www.encodeproject.org/ | Breast, lung, colorectal, hematologic and other cancer cell lines and tissues | Genomics, Epigenomics |
| Synapse (Sage Bionetworks) | https://www.synapse.org/ | Diverse cancers (breast, lung, colorectal, prostate, ovarian, hematologic malignancies, glioma) | Genomics, transcriptomics, proteomics, epigenomics, clinical metadata |
| MetaboLights | https://www.ebi.ac.uk/metabolights/ | Breast, lung, gastric, renal, liver, colorectal, prostate, and multiple myeloma | Metabolomics and associated metadata |
| LINCS (Library of Integrated Network-based Cellular Signatures) | https://lincsproject.org/ | Cancer cell lines (leukemia, lymphoma, breast, lung, colon, prostate, and brain cancer) | Transcriptomics, Proteomics, Metabolomics |
| GDC (Genomic Data Commons) | https://gdc.cancer.gov/ | Multiple cancer types (e.g., colorectal, lung, breast) | Genomics, Transcriptomics, Clinical metadata |
| HuBMAP (Human Biomolecular Atlas Program) | https://hubmapconsortium.org/ | Reference baseline for human tissues (healthy organs; enables cancer atlas comparisons) | Transcriptomics, Proteomics, Spatial Data |
| Cancer Proteome Atlas (TCPA) | https://tcpaportal.org/ | Multiple cancer types (e.g., lung, colorectal, ovarian) | Proteomics, Clinical metadata |
| ImmPort | https://www.immport.org/ | Cancer immunotherapy and tumor microenvironment studies (melanoma, lung, breast, hematologic malignancies) | Genomics, Transcriptomics, Immunomics |
| PRIDE Archive | https://www.ebi.ac.uk/pride/ | Pan-cancer: breast, lung, colorectal, prostate, glioma, leukemia, ovarian, and more. | Proteomics |
| Open Targets Platform | https://www.opentargets.org/ | Multiple cancers (breast, lung, colorectal, prostate, melanoma, leukemia, ovarian, etc.) | Genomics, Clinical |
| BioGRID (Biological General Repository for Interaction Datasets) | https://thebiogrid.org/ | Various (e.g., breast, lung, colorectal, prostate, hematologic, ovarian, etc.) | Protein–protein interactions; genetic interactions (e.g., synthetic lethality) |
| ICGC ARGO (Accelerating Research in Genomic Oncology) | https://www.icgc-argo.org/ | Multiple cancer types (e.g., bladder, breast, colorectal, gastric, head and neck, lung, ovarian) | Genomics, Transcriptomics |
| TCIA (The Cancer Imaging Archive) | https://www.cancerimagingarchive.net/ | Multiple cancer types (e.g., lung, brain, liver) | Pathomics, Radiomics, Genomics, Clinical metadata |
| CPTAC (Clinical Proteomic Tumor Analysis Consortium) | https://hupo.org/Clinical-Proteome-Tumor-Analysis-Consortium-(CPTAC) | Multiple cancer types (e.g., breast, kidney, lung) | Pathomics, Genomics, Proteomics, Clinical metadata |
| Pan-Cancer Atlas | https://gdc.cancer.gov/about-data/publications/pancanatlas | Multiple cancer types (e.g., pancreatic, breast, lung) | Genomics, Transcriptomics, Proteomics |
| BioSamples | https://www.ebi.ac.uk/biosamples/ | Breast, lung, colorectal, prostate, ovarian, melanoma, hematologic malignancies. | Metadata, genomics, transcriptomics, proteomics, metabolomics and epigenomics. |
| Recount2 | https://jhubiostatistics.shinyapps.io/recount/ | Diverse TCGA cancer cohorts (breast, lung, colorectal, prostate, glioma, etc.) | Transcriptomics |
| Xena UCSC | https://xenabrowser.net/ | Multiple cancer types (e.g., lung, breast, pancreatic) | Genomics, Transcriptomics, Epigenomics, Clinical metadata |
| Zenodo | https://zenodo.org/ | Various cancer types (depends on collection; includes multi-omics and imaging cohorts) | Multi-omics (genomics, transcriptomics, proteomics, metabolomics), radiomics/pathomics, and clinical metadata (dataset-specific). |
| Human Tumor Atlas Network (HTAN) | https://humantumoratlas.org/ | Multiple solid tumors and hematologic malignancies (multi-organ tumor atlases) | Single-cell and spatial transcriptomics, proteomics, pathology and advanced imaging, clinical metadata |
| Hest-1K | https://github.com/mahmoodlab/HEST | Cancer cohorts with matched H&E WSIs and spatial omics (tumor-type specific) | Pathomics (WSIs), spatial transcriptomics/proteomics, and associated clinical metadata |

**Table 3**. Conceptual overview of multimodal integration strategies, representative methodological families, and their typical objectives, interpretability characteristics, and input requirement

| Integration Strategy | Representative Method Types | Typical Objectives | Interpretability | Input Requirements |
|---|---|---|---|---|
| Early Fusion | MOFA/MOFA+, iCluster, matrix, tensor, factorization | Unsupervised subtyping, pathway-level biomarker discovery | High (explicit latent factors) | Requires matched modalities with minimal missingness |
| Intermediate Fusion | Attention-based fusion, VAEs, CCA, DCCA, DGCCA | Prognosis, risk stratification, therapy-response prediction | Moderate (latent spaces + attention weights) | Supports heterogeneous modalities; tolerant to partial missingness |
| Late Fusion | Ensemble learning across modality-specific models | Diagnostic classification, workflow augmentation, clinical decision support | Low to moderate (model-level interpretation only) | Robust to missing modalities; flexible combinations |
| Graph-Based Methods | GNNs using PPI, GRN, or biologic interaction networks | Therapeutic vulnerability prediction, mechanism-informed modeling | Moderate to high (node or edge importance, topology) | Requires curated biological networks or inferred graphs |
| Spatial or Non-Spatial Imaging–Omics Fusion | Radiopathomic, pathomic attention-based fusion | Prognosis, biomarker estimation, spatial phenotype discovery | Moderate (spatial attention maps) | Imaging plus optional clinical or molecular data |
| Multimodal Foundation Models | Unified transformer architectures for omics + imaging + text | Diagnosis, prognosis, biomarker discovery | Often complex (attention or saliency improving) | Handles heterogeneous, incomplete, or partially paired inputs |

## 5. Results and Trends from Multimodal Cancer Fusion Studies

Our review and meta-analysis distill the findings from 54 articles that were systematically analyzed to address the objectives of this study. Key characteristics and findings from each study, including methodologies, validation strategies, performance metrics, and data availability, are systematically cataloged to provide a detailed comparative reference (**Table 4**). In addition to summarizing performance, Table 4 explicitly records source code availability, highlighting substantial heterogeneity to the degree to which current multimodal cancer fusion studies enable independent verification and reuse.

### *5.1. Multimodal Data Integration in Cancer Using Machine/Deep Learning*

The earliest multimodal integration pipelines in oncology were built around classical ML models applied to hand-crafted features. In these systems, radiomics, pathomics, or engineered multi-omics signatures were first extracted using predefined feature sets, and then concatenated or combined via early, intermediate, or late fusion strategies (Section 3). Linear models, SVM, RF, XGBoost, and Cox proportional hazards models were widely used for tasks such as survival prediction, recurrence risk stratification, and treatment response classification [106, 113, 114]. These approaches remain attractive in settings with limited sample size or strong prior knowledge because they are relatively data-efficient, straightforward to train and validate, and often more transparent than very deep networks [113].

However, as datasets grew in dimensionality and complexity, particularly when combining multiple omics layers with imaging and clinical variables, the limitations of purely classical models became apparent. Hand-crafted features can miss subtle cross-modal interactions and non-linear structure, and early- or late-fusion schemes built on fixed feature sets often struggle to scale to very high-dimensional spaces without overfitting [115]. These pressures have driven a shift toward deep and hybrid architectures that learn joint representations directly from data, while still using classical ML components (for example, LR, RFs, or Cox models) as prediction heads or baselines [105, 116].

In current state-of-the-art multimodal cancer pipelines, deep models provide the representational backbone, with classical methods retaining value as lightweight, interpretable components and competitive baselines. Broadly, two main research directions have emerged: (i) molecular-level multimodal integration, which fuses multiple omics layers to decode cancer biology, and (ii) fusion of tissue-level imaging with genomic and clinical data to construct a holistic view of the patient.

For molecular multimodal integration, several advanced DL models have demonstrated high performance. Variational autoencoders (VAEs) are prominent, with models like TMO-Net using VAEs with a cross-fusion mechanism to create joint embeddings from CNV, mRNA, mutation, and methylation data for prognostic purposes, achieving F1 scores up to 0.92 and a C-index up to 0.80 [107]. GNNs have emerged as a powerful tool for modeling the complex interactions inherent in biological data. For instance, GNNs have been applied to RNA-Seq, CNA, and methylation data to predict survival across multiple cancer types with a high C-index, and to integrate protein interaction networks with gene expression to predict the effects of kinase inhibitors with an AUC of approximately 0.8 [117]. Ensemble methods, such as the DeepProg framework [118], combine multiple deep learning and machine learning models to integrate gene expression, mutation, and methylation data, yielding robust prognosis prediction with C-indices ranging from 0.68 to 0.80 across various TCGA and GEO datasets.

In the multimodal context, the fusion of imaging data (e.g., WSI, MRI, PET/CT) with other data sources is a key area of innovation. CNNs, often based on architectures like ResNet, remain a foundational component. These are frequently combined with traditional ML classifiers, such as in a model that uses both radiomics and deep features from ResNet50 with a LASSO classifier to achieve an AUC of up to 0.93 for early diagnosis of prostate bone metastasis [105]. Similarly, a ResNet-18 model for pathomics combined with logistic regression for clinical markers reached an AUC of 0.907 for colorectal cancer staging [106]. More advanced architectures are increasingly prevalent. Vision Transformers (ViT) are being integrated with multi-instance learning to combine WSIs and multi-modal data for colorectal cancer prognosis, achieving an AUC greater than 0.85 [119]. Attention mechanisms are central to state-of-the-art fusion models. Pathomic Fusion utilizes a gating-based attention mechanism to integrate histology and genomics, achieving a C-index up to 0.85 for survival prediction [108]. The SAMMS (Spatial Attention-Based Multimodal Survival) model employs multimodal spatial attention to fuse multimodal and histopathology, reaching a C-index of 0.843 for survival prediction in low-grade glioma [120].

**Table 4.** Meta-analysis and data extraction of the selected studies.

| Ref | Year | Methods | Data Types | Integration Type | Validation | Main Results | Aim | Source Code |
|---|---|---|---|---|---|---|---|---|
| [107] | 2024 | TMO-Net:Variational autoencoders with cross-fusion for joint embeddings | CNV, mRNA, mutation, methylation | Molecular multi-omics (genomics, transcriptomics, methylation) | 5-fold CV, CPTAC external | F1:0.75-0.92; C-index:0.60-0.8 | Joint embedding for prognosis | https://github.com/FengAoWang/TMO-Net |
| [105] | 2024 | Radiomics & deep features (ResNet50), LASSO & ML classifiers | WSI, MRI (Pathomics, Radiomics) | Image-based multimodal | Train/val split, 5-fold CV | AUC composite model: 0.85-0.93 | Early diagnosis prostate bone metastasis | N/A |
| [96] | 2024 | Multimodal AI with WSIs & clinical data | WSIs, clinical (PSA, Gleason) | Multimodal clinical/image | Clinical trial validation | HR for DM: 2.33; PCSM: 3.54 | Predict clinical outcomes (PCa) | N/A |
| [101] | 2024 | ML for biochemical recurrence prediction | MRI, WSIs, clinical (PSA, TNM) | Multimodal clinical/image | 363 patients (train/test) | C-index:0.860; AUC:0.911 (3-year BCR) | Predict BCR post-prostatectomy | N/A |
| [106] | 2024 | Pathomics (ResNet-18), Logistic Regression | Pathological slides, clinical markers | Pathological & clinical | Internal dataset (n=267) | AUC combined:0.907 train,0.814 test | Colorectal cancer staging | N/A |
| [121] | 2024 | AI-LLMs integrated with CNN/Transformer | Clinical data, CT images | Text & image fusion | 163 patients, 64 tests | Extraction accuracy:87-97%; AUC:0.89 max | Survival prediction bladder cancer | N/A |
| [122] | 2023 | ML for cis-regulatory elements | Multi-omics (single cell) | Multi-omics molecular | Cross-species validation | Accuracy:0.65-0.90; correlation:0.8-0.88 | Predict epigenetic conservation | https://github.com/ejarmand/comparative_epigenomic_motor_cortex |
| [123] | 2023 | Geometric Graph Neural Networks (GGNN) | RNA-Seq, CNA, methylation | Multi-omics molecular | Multiple cancers (TCGA) | High C-index (except colon cancer) | Survival prediction | (https://github.com/MSK-MOI/GGNN |
| [111] | 2023 | CNNs for PET/CT quality assessment | PET/CT imaging | Imaging | Internal dataset (173 patients) | Kappa:0.78-0.80; ICC:>0.75 | Automated PET/CT quality assessment | N/A |
| [124] | 2023 | Radiomics via SDCT | Imaging (SDCT sequences) | Imaging multimodal | Internal dataset (176 patients) | AUC combined:0.961 train,0.944 test | Predict lung adenocarcinoma invasiveness | N/A |
| [125] | 2023 | Contrastive Learning (PLIP) | Pathology images/text | Visual-language multimodal | Zero-shot evaluation, fine-tuned | F1:0.856-0.927; Recall@10:0.557 | Visual-language foundation pathology | https://github.com/PathologyFoundation/plip |
| [120] | 2023 | Multimodal spatial attention (SAMMS) | Multi-omics & histopathological | Spatial attention multimodal | 5-fold CV (TCGA) | C-index:0.843; AUC:0.782 (LGG) | Cancer survival prediction | N/A |
| [117] | 2022 | GNNs for therapeutic effects prediction | Protein interactions, gene expression | Molecular multi-omics | CCLE, LINCS | AUC: ~0.8 | Predict kinase inhibitors effect | https://github.com/pu-limeng/CancerOmicsNet |
| [119] | 2022 | Vision Transformer & multi-instance learning | WSIs, multi-omics | Image & molecular multimodal | Independent cohorts & TCGA | AUC:>0.85; F1:>0.80 | Prognosis colorectal cancer | https://github.com/pu-limeng/CancerOmicsNet. |
| [113] | 2021 | MRI radiomics & SVM classifiers | MRI, clinical-pathologic | Imaging & clinical | Multicenter dataset | AUC:0.90-0.93 | ALN metastasis prediction | https://github.com/ZifanHe/Ebiomedicine.git |
| [109] | 2022 | Radiopathomics (RAPIDS) | MRI, WSIs | Image multimodal | Multicenter observational | AUC:0.81-0.86 | Predict chemotherapy response | Radiopathomics (RAPIDS) |
| [126] | 2021 | CNN (VGG16, ResNet50) | Colposcope images | Image multimodal | Internal dataset | Accuracy:86.3% | Cervical lesion classification | N/A |
| [127] | 2021 | U-Net with multimodal attention | PET/CT images | Imaging multimodal | 5-fold CV | Dice:71.44-62.26 | Tumor segmentation | N/A |
| [112] | 2021 | CNN & Word2Vec fusion | Endoscopy images & text | Multimodal fusion | Dataset split | Accuracy:95.3%; sensitivity:90.2% | UGI cancer screening | https://github.com/netfly-machine/SCNET |
| [128] | 2021 | Richer Fusion Network | Pathology images & EMR | Multimodal fusion | Dataset split | Accuracy:92.9% | Breast cancer classification | N/A |
| [110] | 2020 | MSReg with 3D ResNeXt | MRI & Ultrasound | Image registration | NIH clinical trial (n=679) | Error reduced up to 62% | Prostate cancer biopsy guidance | N/A |
| [116] | 2020 | EPLA with MIL & ResNet-18 | Histopathology, multi-omics | Multimodal fusion | TCGA, Asian-CRC | AUC:0.8848-0.8504 | Predict MSI colorectal cancer | https://github.com/yfzon/EPLA |
| [129] | 2023 | Transformer-based PathOmics | WSI, mRNA, CNV, DNA methylation | Multimodal embeddings | TCGA COAD, TCGA-READ | AUC:0.56-0.75 | Survival prediction in colon cancer | https://github.com/Cas-sie07/PathOmics |
| [130] | 2021 | ML classifiers for cfDNA analysis | cfDNA liquid biopsy data | Integrative analysis | 10-fold CV | AUC:0.76-0.97 | Pediatric sarcoma diagnosis | https://medical-epige-nomics.org/pa-pers/peneder2020_f17c4e3befc643ffbb31e69f43630748/ |
| [100] | 2021 | DL (SResCNN) for PD-L1 prediction | PET/CT, clinical data | Multimodal fusion | Internal, external cohorts | AUC:0.82-0.89; accuracy:77.7%-81.7% | PD-L1 prediction NSCLC | N/A |

| [114] | 2021 | XGBoost-based radiomics | PET/CT imaging | Imaging multimodal | Internal dataset | Sensitivity:90.9%; specificity:71.4%; accuracy:80% | Predict ALN metastasis (IDC) | N/A |
|---|---|---|---|---|---|---|---|---|
| [118] | 2021 | Ensemble DL & ML (DeepProg) | Gene expression, mutation, methylation | Multi-omics integration | TCGA, GEO datasets | C-index:0.68-0.80 | Prognosis prediction | https://github.com/lana-garmire/DeepProg |
| [131] | 2021 | Ensemble random forest radiomics | PET/MRI imaging (Ga-PSMA-11, ADC, T2w) | Imaging multimodal | Internal dataset (52 patients) | AUC:0.86-0.94; accuracy:81%-91% | Risk assessment prostate cancer | N/A |
| [130] | 2021 | LIQUORICE ML for cfDNA patterns. | Genomic, epigenomic (cfDNA) | Integrative analysis | ENCODE, TCGA | AUC:0.97, Sens:85% @ 100% Spec | cfDNA-based pediatric cancer detection (EwS/sarcoma) | https://liquorice.readthedocs.io/en/latest/intro.html?utm_source=chatgpt.com |
| [132] | 2020 | IOUC-3DFCNN with Haar-like fusion | MRI (T1, T1GD, T2, Flair) | Multimodal auto-context | BRATS 2017 & 2013 datasets | DICE:0.70-0.89; Recall:0.79-0.91 | Brain tumor segmentation (Gliomas) | N/A |
| [133] | 2022 | CPH with L1/L2 regularization | WSI, CT, clinicogenomics | Late multimodal fusion | Train-test split (444 patients) | Significant correlation with GVHD, OS:81%, PFS:76% | Predict GVHD after bone marrow transplant | https://github.com/kmboehm/onco-fusion |
| [134] | 2022 | CART analysis on immunophenotypic, proteomic, clinical data | Immunophenotypic, proteomic, clinical | Multimodal integration | Two trials (145 patients) | Significant correlation with GVHD, OS:81%, PFS:76% | Predict GVHD after bone marrow transplant | N/A |
| [108] | 2020 | Pathomic Fusion with gating-attention | Histology, genomics (CNV, RNA-seq) | Multimodal fusion | 15-fold CV | Accuracy:85-95%, ROC-AUC:0.92-0.98, C-Index:0.72-0.85. | Survival outcome prediction | https://github.com/mahmoodlab/PathomicFusion |
| [135] | 2020 | NMF clustering, DGE, pathway & co-occurrence analysis | Transcriptomic, somatic genomic | Multi-omics clustering | Internal validation, TCGA | Identified prognostic mutations and therapy sensitivity | Subtyping RCC and therapy response | N/A |

Furthermore, the field is beginning to leverage large language models (LLMs), with one study integrating LLMs with CNNs/Transformers to fuse clinical text and CT images for bladder cancer survival prediction, demonstrating high data extraction accuracy (87-97%) and an AUC of 0.89. Contrastive learning approaches, such as in the Pathology Language-Image Pre-training (PLIP) model, are creating powerful visual-language FMs for pathology, achieving high F1 scores (up to 0.927) in zero-shot evaluations [125].

*5.2. Reproducibility and Proprietary Barriers*

A striking pattern emerging from Table 4 is that only a subset of high-performing models provide open-source code or fully accessible pipelines, and many rely on institutional or proprietary datasets that cannot be freely shared. This combination of closed implementations and restricted data access poses a significant barrier to reproducibility, independent benchmarking, and fair comparison across methods. In some cases, reported performance cannot be rigorously verified outside the originating center, limiting the community's ability to assess robustness, detect overfitting, or adapt the models to new populations. For multimodal integration and FMs in particular, where training often requires large-scale, heterogeneous cohorts, these constraints are especially problematic: models trained on proprietary imaging–omics–clinical repositories may encode site-specific biases that are difficult to detect without external replication. More broadly, the lack of standardized reporting on data preprocessing, integration strategies, and cross-site validation further complicates reuse. Going forward, concrete steps such as releasing source code and pretrained weights whenever possible, prioritizing evaluation on public multimodal benchmarks, and adopting transparent reporting checklists (e.g., for data splits, harmonization, and domain shifts) will be crucial to strengthen reproducibility in this field. For settings where data cannot be shared directly, approaches such as federated or privacy-preserving training, shared synthetic benchmarks, and detailed protocol publication offer pragmatic pathways to mitigate the limitations imposed by proprietary datasets and methods [136]. In this review, we explicitly note where reported gains rely on single-center cohorts, lack external validation, or conflict with findings from related studies, and we qualify our interpretation of such results accordingly in Table 4 and Section 5.3. These reproducibility and evidence-quality limitations are further synthesized in Section 6.1 as key gaps that must be addressed before many of the surveyed approaches can be considered ready for routine clinical deployment.

*5.3. Translating Technological Advances into Clinical Outcomes*

In Table 4, multimodal models are evaluated on a diverse set of endpoints that are directly tied to patient care. Many works focus on survival-related metrics (e.g., overall survival, progression-free survival, biochemical recurrence, or graft-versus-host disease risk), using C-indices [123] or hazard ratios to quantify prognostic value [108, 120]. For example, multi-omics survival models such as GGNN [123], DeepProg [118], SAMMS [120], and transplant-focused fusion models [133, 134], as well as multimodal colon cancer survival frameworks like PathOmics [129] and Pathomic Fusion [108]. Other studies target early detection or staging, including nodal and distant metastasis prediction and tumor invasiveness from radiology–clinical or pathomics–clinical fusion [96, 101, 106, 114].Others target early detection or staging (e.g., nodal metastasis prediction, lesion classification, or tumor invasiveness), treatment response (e.g., chemotherapy efficacy, immunotherapy benefit, kinase inhibitor response [117]), or biomarker surrogates such as PD-L1 [100] status and microsatellite instability [116]. A smaller but important subset addresses workflow-centric tasks, including automated PET/CT quality assessment [111], multimodal MRI–ultrasound registration for biopsy guidance [110], and segmentation of primary and nodal disease from PET/CT or brain MRI using multimodal CNN and U-Net architectures [127, 132, 137]. Taken together, these endpoints illustrate that recent technological advances in multimodal integration and FMs are not operating in a vacuum, but are increasingly aligned with clinically meaningful tasks. For example, models that combine radiology, pathology, and multi-omics to refine risk stratification could enable more granular decisions about adjuvant therapy, surveillance intensity, or trial eligibility [96, 101, 106, 108, 114, 118, 120, 123, 133]. Early metastasis prediction or accurate nodal staging from integrated imaging–clinical models has the potential to reduce unnecessary procedures, prioritize high-risk patients for aggressive treatment, and avoid overtreatment in low-risk groups [113, 131]. Similarly, multimodal prediction of treatment response or immune-related biomarkers (such as PD-L1 or MSI) could inform therapy selection, sparing patients from ineffective regimens and associated toxicities [100, 109, 116, 117]. Workflow-oriented models, including those for image quality control, registration, and auto-segmentation, primarily target healthcare efficiency by reducing manual workload and variability, shortening time-to-report, and improving consistency across institutions [110, 111, 127, 132].

At the same time, most studies in our survey remain retrospective, single- or limited-center analyses, and only a minority are embedded in prospective trials or decision-analytic frameworks. Health-economic endpoints such as cost-effectiveness, resource utilization, or impact on waiting times are rarely quantified. As a result, the link from improved predictive performance to tangible clinical benefit is often inferred rather than directly demonstrated. Bridging this gap will require prospective impact evaluations, integration of multimodal FMs into clinical decision-support tools, and formal assessment of how their use affects survival, toxicity burden, time-to-diagnosis, and healthcare efficiency at the system level. These needs are echoed and further elaborated in the future directions section (Section 6.2).

### *5.4. State-of-the-Art Foundation Models*

FMs represent a paradigm shift in computational oncology, moving beyond task-specific supervised learning toward architectures that learn generalized representations of complex biological data. By leveraging self-supervision on vast, often unlabeled datasets, these models extract deep, transferable features from individual modalities such as omics, histopathology, and radiology. Building on the classical ML and deep/hybrid architectures discussed in Section 5.1 and the integration patterns outlined in Section 3, this section has two goals: (i) to summarize the current landscape of medical FMs across modalities, and (ii) to highlight emerging algorithmic innovations and domain-specific design choices that distinguish oncologic FMs from generic vision or language models. **Figure 6** provides a high-level taxonomy of FMs in multimodal cancer research, spanning (i) unimodal omics, (ii) pathology, (iii) radiology, and (iv) integrative cross-domain models. In the following, Section 5.4.1 offers a model-level survey (complemented by **Table 5**), while Section 5.4.2 focuses on architectural and algorithmic considerations specific to the medical domain.

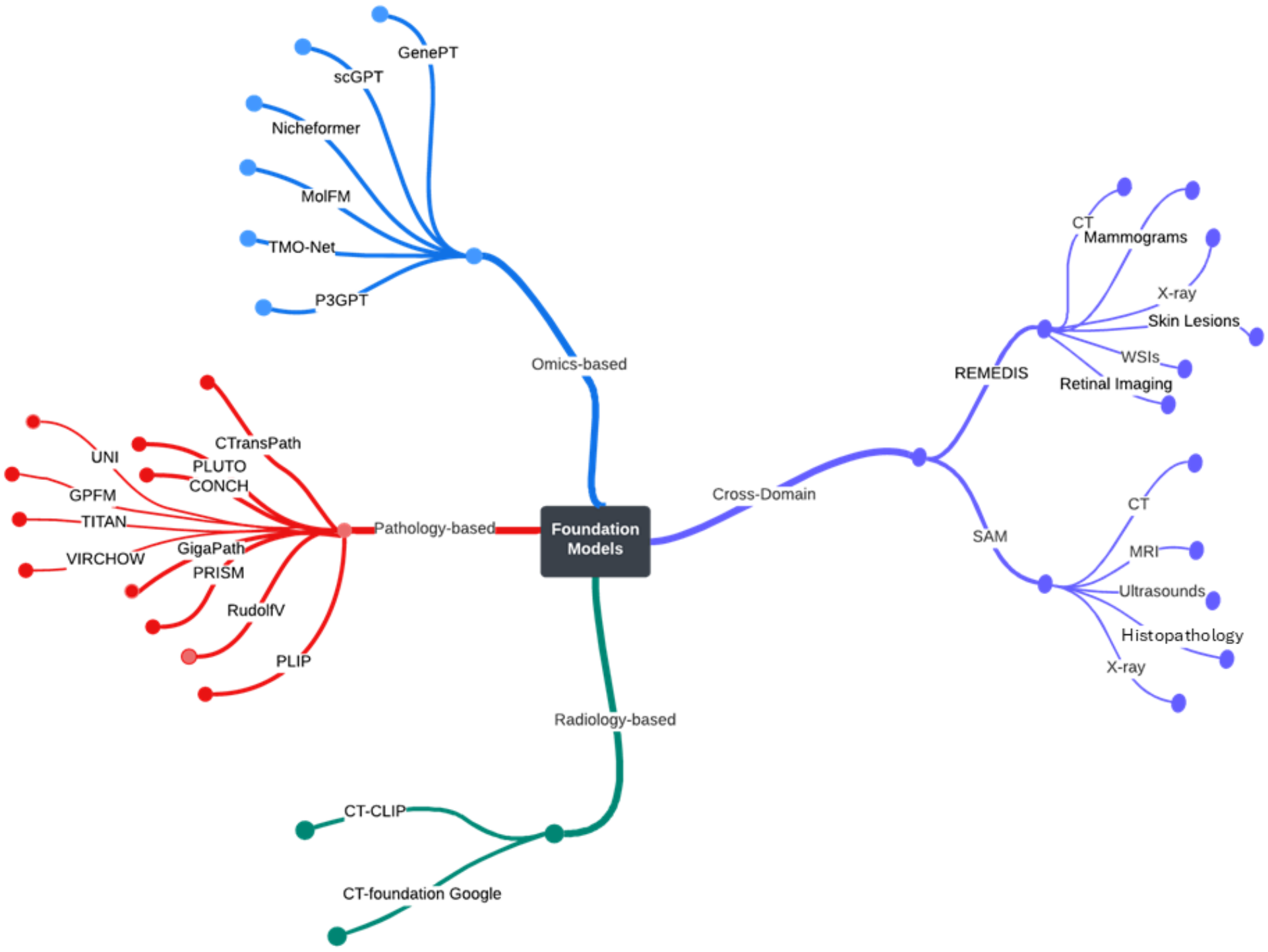

**Figure 6.** A proposed taxonomy for foundation models in multimodal cancer research. The framework delineates four key domains: (i) unimodal models for omics, (ii) pathology, and (iii) radiology, and (iv) integrative models designed for cross-domain data.

#### *5.4.1. Taxonomy of Cancer Foundation Models Across Modalities*

FMs are designed to integrate diverse data layers to better characterize cancer heterogeneity, therapeutic vulnerabilities, and disease progression. **Table 5** summarizes representative models across four domains—omics-based, pathology-based, radiology-based, and cross-domain multimodal systems— detailing for each the primary objective, training modalities, cohorts, training strategy, and downstream tasks.

*A. Omics-based and molecular FMs.*

On the molecular side, several FMs learn high-capacity representations from large-scale multi-omics or single-cell datasets. TMO-Net, for example, adopts a multi-task learning framework including gene mutation, DNA methylation, transcriptomics, and copy number variations (CNVs) to reveal complex patterns in a regularity network [107]. This network could be fine-tuned for downstream tasks even with incomplete data modalities. Similarly, scGPT adopts a pretrained transformer for single-cell biology [44]. The performance of scGPT model is validated for several downstream tasks, such as single-nucleus sequencing for pancreatic ductal adenocarcinoma (PDAC) [138], cell type annotation [139], to assess functional characterization of the immune system [3], cross-tissue single-cell annotation [140], generative modeling of omics data [141], and generation of single-cell data [142]. Nicheformer [143] is designed to analyze single-cell and spatial transcriptomics (ST) data. This model is pretrained on 57 million dissociated and 53 million spatially resolved cells across 73 tissues on cellular reconstruction. The nicheformer is applied in several downstream analysis tasks, such as spatial label prediction, prediction of the effect of unseen single and double perturbations, and single-cell omics [144]. MolFM jointly learns from molecular structures, biomedical texts, and knowledge graphs and is applied to cross-modal retrieval, molecule captioning, and text-based molecule generation tasks [43]. GenePT [145] represents genes and cells by leveraging LLM embeddings using text descriptions from the National Center for Biotechnology Information (NCBI). GenePT is applied to gene functionality classification, gene property prediction, gene-gene interaction (GGI), protein-protein interaction (PPI) prediction, and gene perturbation prediction for downstream tasks [146], while Precious3-GPT adopts a multimodal approach to combine textual, tabular, and knowledge graph representations of biological experiments to study the biological intersections of aging and cancer and evaluate for downstream tasks such as age prediction, drug sensitivity prediction, and gene annotation [147]. Collectively, FMs, including TMO-Net [107], P3GPT [147], Path-GPTOmic [148], MolFM [43], and Nicheformer [143], have significantly advanced the integration ability of diverse molecular data and offered holistic insights into the complex biological. However, while these models unravel the molecular intricacies of cancer, they lack the spatial and structural information that imaging modalities provide.

*B. Pathology and Radiology FMs.*

Imaging FMs capture the structural and morphological hallmarks of cancer. Histopathology-focused models such as UNI [149], PLUTO [150], GPFM [151], RudolfV [151], and CTransPath [152] are trained on large collections of WSIs to produce robust slide- or patch-level embeddings. Others, including PRISM [153] and CONCH [154], adopt vision–language objectives to align WSIs with clinical reports or captions. Parallel efforts in radiology have yielded dedicated FMs for CT and other imaging modalities. For example, SAM-Med [56], CT-CLIP [155], and Google's CT-Foundation model [156], which transform volumetric CT data into compact, information-rich embeddings suitable for downstream tasks such as organ segmentation, lesion detection, and cancer risk prediction. These imaging FMs accelerate diagnostics and image-based biomarker discovery by providing rich spatial representations of tumor morphology and microenvironment. Within computational pathology, we note that several domain-specific reviews and toolboxes (e.g., large-scale surveys of AI in digital pathology and open-source frameworks such as TIAToolbox for end-to-end WSI analysis) already cover digital pathology workflows in depth [157]; here we focus specifically on their role as FMs within multimodal oncology pipelines.

*C. Cross-domain and integrative FMs.*

A smaller but rapidly growing class of models explicitly integrates imaging, molecular, and clinical data within a unified FM. Examples include multimodal architectures such as Path-GPTOmic [148] and P3GPT-like systems, which couple cross-attention or contrastive objectives across WSIs, omics, and textual/clinical inputs. These integrative FMs typically require carefully curated, paired datasets but hold the greatest promise for capturing cross-scale biology and supporting diverse tasks—ranging from survival prediction and treatment response modeling to zero-shot biomarker inference and clinical report generation.

From a comparative standpoint, these model families occupy different positions along the trade-off space between model complexity, data requirements, and performance. Unimodal omics FMs (e.g., scGPT [44], Nicheformer [143], MolFM [43], GenePT, Precious3-GPT [147]) operate on very large but relatively well-structured datasets and provide strong transferability within the molecular domain, at the cost of substantial pretraining compute. Histopathology and radiology FMs must process giga-pixel WSIs or 3D CT volumes, driving architectural complexity (hierarchical ViTs, MIL-based pipelines) and memory footprint, but yielding highly expressive spatial representations. Integrative multimodal FMs sit at the most demanding end of this spectrum, requiring paired imaging–omics–clinical cohorts and more elaborate fusion modules, yet offering the broadest potential clinical utility [115]. Importantly, in low-data regimes or narrowly defined tasks, simpler classical ML models or shallow DL architectures remain competitive due to lower data requirements, easier calibration, and more straightforward validation.

**Table 5.** Summary of Cancer Research Studies Employing **Foundation Models** in Oncology.

| Omics-based Foundation Models | | | | | | |
|---|---|---|---|---|---|---|
| **Foundation Model** | **Year** | **Objective/Purpose** | **Training Modality** | **Training Cohorts** | **Training Method** | **Downstream Tasks** |
| GET [158] | 2025 | Models regulatory grammar across human cell types | Chromatin accessibility + DNA sequence | 213 human fetal and adult cell types | Transformer-based FM using chromatin accessibility data and sequence information | Gene expression, TF interactions, regulatory element and response prediction |
| OmniCLIP [159] | 2025 | Visual omics FM linking WSIs with spatial transcripts via contrastive learning | WSIs + STs | 2.2 million paired tissue images from 1,007 samples across 32 organs using 10x Visium spatial data | CLIP-based FM using contrastive learning, by using transcriptomic data into sentences with a dual encoder architecture | Tissue alignment, annotation, decomposition, retrieval, spatial gene prediction. |
| scGPT [44] | 2024 | Distills critical biological insights concerning genes and cells | Multi-Omics | Single-Cell Sequencing Data over 33 million cells | A generative pretrained transformer | Cell type annotation, multi-omics integration, response prediction, gene network inference |
| Nicheformer [143] | 2024 | Cell embeddings reflect local microenvironment context. | Single-Cell STs + Spatially Resolved Transcriptomics | 110M scRNA-seq + spatial cells (MERFISH, Xenium, CosMx, ISS) | Transformer-based models use gene-rank tokenization with a 12-layer, 16-head transformer. | Spatial density prediction, spatial label prediction, cross-modal transfer |
| MolFM [43] | 2023 | Multimodal FM learning from molecular structures, texts, and knowledge graphs**.** | Molecular structures + knowledge graphs + and biomedical texts | 5k structures, 37M text paragraphs, 49k-entity KG (3.2M relations) | Cross-modal attention aligning molecular modalities while maintaining global knowledge. | Cross-modal retrieval, molecule captioning, text-based molecule generation, and molecular property prediction |
| GenePT [145] | 2023 | FM representing genes and cells using LLM-derived embeddings | NCBI gene database. | Use GPT-3.5 to create embeddings from the NCBI text descriptions of genes. | LLM ChatGPT (GPT-3.5) | Gene functionality classification, gene property prediction, GGI and PPI prediction, Cell Type Annotation |
| TMO-Net [107] | 2024 | Multi-omics integration for pan-cancer with missing-modality support. | Gene Mutation, mRNA Expression, CNV, DNA Methylation Profiles data | TCGA which includes 8,174 samples | multi-task pre-training with cross-fusion module and self-supervised contrastive learning | cancer subtyping, metastasis prediction, drug response prediction, prognosis prediction |
| Precious3-GPT (P3GPT) [147] | 2024 | Multimodal transformer FM for aging and drug discovery using multi-species multi-omics data. | RNA-seq (GTEx, ARCHS4, LINCS) + DNA methylation and proteomics datasets. | 1.2M samples across 63k+ biological entities (multi-species) | Prompt-driven multimodal transformers (omics + text) | Age prediction, Target identification, drug sensitivity prediction, gene annotation |

<table>
<tr><td>DNABERT-2 [160]</td><td>2023</td><td>Multi-species genomics FM overcoming k-mer tokenization limits</td><td>DNA Sequences</td><td>Multi-species genome dataset combining 36 datasets across 9 tasks (70–10,000 bp)</td><td>Uses Byte Pair Encoding (BPE) + transformer for long multi-species genomes</td><td>Genomic variant prediction, cancer mutation assessment, genomics analysis, cancer genomics applications</td></tr>
<tr><td colspan="7"><b>Pathology-based Foundation Models</b></td></tr>
<tr><td><b>Foundation Model</b></td><td><b>Year</b></td><td><b>Objective/Purpose</b></td><td><b>Training Modality</b></td><td><b>Training Cohorts</b></td><td><b>Training Method</b></td><td><b>Downstream Tasks</b></td></tr>
<tr><td>MUSK [161]</td><td>2025</td><td>Multimodal transformers for predicting cancer by integrating pathology images with clinical text data through cross modal learning</td><td>WSIs + Text</td><td>50 million pathology images paired with 1 billion pathology related text descriptions from multiple cancer centers and clinical databases.</td><td>Leverages unified mask modeling approach to handle unpaired multimodal data with cross modal attention, contrastive learning between image and text modality</td><td>Disease level survival prediction, cancer subtyping, biomarker identification, treatment response prediction</td></tr>
<tr><td>Prov-GigaPath [162]</td><td>2024</td><td>Clinical grade foundation model for digital pathology for cancer diagnosis and prognosis</td><td>WSIs</td><td>1.3 billion 256 × 256 pathology image tiles in 171,189 whole slides 30,000 patients covering 31 major tissue types</td><td>Tile level training pretraining using DINOv2 [163], slide-level training using masked level autoencoder, vision language pretraining with clinical reports</td><td>Biomarker prediction, mutation status prediction, survival analysis, cancer subtyping</td></tr>
<tr><td>PathCHAT [164]</td><td>2024</td><td>Vision language AI assistant for human pathology - multimodal generative AI copilot</td><td>WSIs + Text</td><td>456,000 different visual language instructions consisting of 999,202 question and answers</td><td>Pretrained large language model and fine-tuned on visual language instructions</td><td>Pathological query assistance, diagnostic support, morphological analysis, differential diagnosis</td></tr>
<tr><td>BEPH [165]</td><td>2025</td><td>Uses self-supervised learning (SSL) for generalizable cancer diagnosis and survival prediction</td><td>WSIs</td><td>11 million histopathological images from multiple cancer subtypes</td><td>SSL for learning meaningful representations from histopathological images</td><td>Patch-level cancer prediction, cancer classification, survival prediction for cancer types</td></tr>
<tr><td>CHIEF [166]</td><td>2024</td><td>Uses weakly-supervised machine learning to extract pathology imaging features</td><td>WSIs</td><td>60,530 WSIs consisting of 19 anatomical sites</td><td>Unsupervised pretraining for tile level feature identification and weakly supervised pretraining for whole slide pattern recognition</td><td>Cancer cell detection, prognostic prediction, tumor identification, genomic prediction</td></tr>
<tr><td>PLUTO [150]</td><td>2024</td><td>The generation of task-agnostic embeddings.</td><td>WSIs</td><td>Pretrained on a dataset of 195 million tiles from 158, 852 WSIs.</td><td>DINO [167] + iBOT SSL with MAE and Fourier losses to preserve high-frequency detail</td><td>Instance segmentation, tile classification, slide-level class, and survival analysis.</td></tr>
<tr><td>UNI [149]</td><td>2024</td><td>Interpreting pathology images; surgical pathology</td><td>WSIs</td><td>100 million images from over 100,000 diagnostic WSIs across 20 major sub-types</td><td>SSL and supervised fine-tuning</td><td>Slide classification, ROI classification and retrieval, cell type segmentation, and slide prototyping.</td></tr>
</table>

| | | | | | | |
|---|---|---|---|---|---|---|
| GPFM [151] | 2024 | Improve the generalization capability of pathology FMs across a wide range of clinical tasks. | WSIs | 86,000 WSIs across 34 major tissue types | Expert and self-knowledge Distillation | Slide Classification, ROI analysis, Image retrieval, Pathological Images VQA, and Report generation. |
| TITAN [168] | 2024 | Incorporates multimodal learning by aligning visual features of pathology slides with synthetic captions and clinical reports. | WSIs + Text + synthetic Region-of-Interest (ROI) Captions | 335,645 WSIs with 423,122 synthetic captions at the region-of-interest, and 82,862 clinical pathology reports. | Three-stage pretraining: SSL on WSIs, then alignment with captions and clinical reports. | Survival Prediction, Morphological Classification, Molecular Classification, Grading Tasks, Cross-Modal Retrieval, Pathology Report Generation |
| VIRCHOW [46] | 2024 | To develop clinical-grade FM for digital pathology by enabling the detection of both common and rare cancers | WSI | 1.5 million H&E-stained WSIs from around 100,000 patients | The model was trained using the DINO v2 framework [163], a self-supervised multi-view student-teacher approach | Pan-Cancer Detection and subtyping, tile-level prediction, survival analysis, |
| GigaPath [169] | 2024 | GigaPath is to develop to generalize effectively across diverse real-world clinical settings | WSIs | 1.3 billion 256 × 256 image tiles from 171,189 whole slides. | Three-stage training, which includes tile-level pretraining: Using DINOv2 [163], slide-level training using a masked autoencoder and LongNet, and VLM pretraining. | 17 pathomics tasks, Mutation Prediction, and zero-shot cancer subtyping |
| RudolfV [170] | 2024 | To develop a pathology foundation model of diverse pathologist-guided curation. | WSIs | 134,000 slides from 34,103 cases grouped by semantic similarity, and clustered tiles to ensure meaningful pattern learning. | Trained using DINOv2 [163] for SSL | Slide classification, reference search, IHC Biomarker Scoring, Histological and Molecular Prediction |
| CONCH [154] | 2024 | To develop a visual-language FM for computational pathology that overcomes limitations such as label scarcity and task-specific training. | WSIs +Text | Trained using over 1.17 million image–caption pairs sourced from various educational and clinical materials | Visual-Language pretraining using a contrastive learning approach based on the CoCa (Contrastive Captioners) framework [171]. | Slide-level classification, ROI classification, cross modality retrieval |
| PRISM [153] | 2024 | To efficiently aggregate information from image tiles and clinical reports to create holistic | WSIs + Text | 587,000 whole slide images (WSIs) paired with 195,000 associated clinical reports. | Adopts the CoCa framework, which comprises a vision encoder, Language Decoder (BioGPT), and Contrastive Learning. | Subtyping, biomarker prediction, clinical report generation, zero shot prediction |
| | | representations of WSI, combining | | | | |
| PLIP [125] | 2023 | Crowd-sourced image–text data used to train a generalizable FM. | WSIs + Text | 208,414 pathology images paired with natural language descriptions. | Multimodal AI (language & image pretraining) | Slide classification, Image–Text Retrieval and Segmentation/Captioning |
| CTransPath [152] | 2022 | To develop a self-supervised feature extractor specifically designed for histopathological image classification. | WSIs | 15 million tiles derived from WSIs over 25 anatomical sites and cover 32 different cancer subtypes. | Semantically Relevant Contrastive Learning (SRCL) combined with CNN backbone. | Slide and tile classification, Patch retrieval, Gland segmentation, mitosis detection |
| **Radiology-based Foundation Models** | | | | | | |
| **Foundation Model** | **Year** | **Objective/Purpose** | **Training Modality** | **Training Cohorts** | **Training Method** | **Downstream Tasks** |
| MEDFORM [172] | 2025 | Medical FM that uses CT with clinical variables for multimodal cancer analysis and risk prediction | CT + Clinical Data | Pretrained on CT slices from lung (141k), breast (8k), and colorectal cancer (10k) | MIL with SimCLR SSL for CT features, followed by CT–clinical alignment via contrastive learning | Cancer risk prediction, treatment response assessment and monitoring, multimodal cancer staging classification |
| CT-FM [173] | 2025 | 3D image-based FM for volumetric CT understanding for clinical cancer imaging | CT | 148,000 CT scans consisting of anatomical regions, cancer types, and imaging protocols | Uses label level contrastive learning for 3D volumetric data processing, stability optimization | Whole body organ segmentation, tumor segmentation and detection, cancer detection |
| MEDSAM2 [55] | 2024 | Medical segmentation model with memory attention formedical imaging and video analysis | CT + MRI + Ultrasound + multi-dimensional imaging | Medical dataset with over 455,000 3D image-mask pairs and 76,000 frames | Memory-attention blocks for processing sequential medical images and 3D volumes | 3D organ segmentation, temporal medical image analysis, real-time ultrasound segmentation, cardiac imaging |
| M3FM [174] | 2025 | Medical multimodal FM for cancer analysis and risk prediction | CT + Clinical data | 163,725 chest CT with 49 different clinical data types | Utilizes multimodal question answering with distributed parallel training along with cross modal attention mechanism. | Lung cancer risk prediction with cancer assessment across imaging and clinical modalities |
| CT-CLIP [155] | 2024 | Generalist FM for 3D CT | CT + Text | 25,692 non-contrast 3D chest CT scans from CT-RATE, approximately 14.3 million 2D slices. | Contrastive learning approach with zero-shot multi-abnormality detection, and fine-tuning, using ClassFine and VocabFine | Multi-abnormality detection, case retrieval, multimodal interaction with CT-CHAT |
| CT-foundation Google [156] | 2024 | Converts CT volumes into rich embeddings for | CT + radiology report | 527k CT studies with reports from 430k patients | 2D CoCa model [175] extended to VideoCoCa [171], trained on 500k+ CT–report pairs | Hemorrhage, calcifications, lung cancer, abdominal |
| | | efficient task-specific modeling | | across three U.S. hospital regions | | lesions, urolithiasis, aneurysm detection |
| **Cross-Domain Foundation Models** | | | | | | |
| **Foundation Model** | **Year** | **Objective/Purpose** | **Training Modality** | **Training Cohorts** | **Training Method** | **Downstream Tasks** |
| BiomedParse [174] | 2025 | FM for biomedical image parsing, joint object detection, and segmentation across modalities | Image + Text | 6M+ image-mask-label sets from 1M+ images across 64 major biomedical object types | Joint pretraining using GPT-4 synthesized data from 45 existing segmentation datasets | Biomedical object segmentation, detection, recognition across 9 modalities |
| MedImageInsight [176] | 2024 | medical imaging embedding model for general domain medical imaging | Image + Text | Medical images across X-Ray, CT, MRI, dermoscopy, OCT, fundus photography, ultrasound, histopathology, mammography | Contrastive learning with medical image text pairs | Medical image classification, image search, report generation, disease classification |
| SeNMo [54] | 2024 | Self-Normalizing Foundation Model to address multiple omics modalities and clinical data types for cancer analysis | Multi-Omics + Clinical Data | The training of multi omics data includes gene expression, DNA methylation, miRNA expression, DNA mutations, protein expression modalities, and clinical data. | Self-normalizing approach for cross dataset generalization, multi-omics integration, domain adaptation techniques for handling institutional and technical batch effects | Overall Survival Prediction, Primary Cancer Type Classification, Tertiary Lymphoid Structures, Ratio Prediction |
| SAM [56] | 2024 | Adapts SAM for unified segmentation across multimodal medical imaging | CT, MRI, ultrasound, X-ray, colonoscopy, WSIs, electron microscopy, fundus imaging | 1,050,000 2D medical images with Approximately 6,033,000 segmentation masks | Pretrained on 11M images with 1B masks (ViT), then fine-tuned on COSMOS-1050K | Organ segmentation, lesion segmentation, histopathological segmentation, everything mode evaluation |
| REMEDIS [57] | 2023 | Generalizable FMs through combined supervised and SSL multimodal pretraining. | ImageNet-21K, CT scans, Chest X-rays, WSIs, Mammograms, Skin Lesions, Retinal Imaging | Several data sources are used to collect multimodal data | Supervised pre-training and SSL. | Thoracic/breast cancer, tumors, segmentation, retinal and skin lesion detection |

### 5.4.2. Algorithmic Innovations and Domain-Specific Model Design

While Section 5.4.1 focuses on what cancer FMs currently exist, an equally important question is how their architectures must be adapted for oncology, beyond simply applying generic vision or language backbones to medical datasets. In practice, contemporary oncologic FMs span a spectrum of modelling intent from task- and domain-specific encoders (e.g., pathology-only or CT-only models optimized for a restricted set of endpoints) to broad, general-purpose representations that are pretrained with generic objectives and then adapted to multiple diagnostic, prognostic, and treatment-related tasks. This intent is layered on top of the integration patterns described in Section 3 (e.g., early vs. intermediate vs. late fusion; attention-, graph-, or correlation-based strategies), rather than replacing them. **Figure 7** summarizes these design elements across backbone choice, self-supervised pretraining, and downstream adaptation.

Several algorithmic themes are emerging as particularly domain-critical for medical FMs. First, biology-aware architectures increasingly embed prior biological structure—for example, gene-rank tokenizations and pathway-aware attention in omics models, or hierarchical encoders that mirror tissue, organ, and patient-level organization in imaging FMs. These designs move beyond generic spatial or textual encoders toward networks that better reflect biological hierarchies and dependencies. Second, uncertainty quantification and calibration are becoming central requirements for clinical deployment: Bayesian layers, deep ensembles, temperature scaling, and conformal prediction are being coupled to FM backbones to provide calibrated probabilities, uncertainty estimates, and risk strata that can be meaningfully incorporated into clinical decision-making and regulatory assessment [177].

Third, domain adaptation and robustness are essential because medical data exhibit strong site-, vendor-, and protocol-specific shifts. Domain-aware FMs therefore incorporate components such as conditional normalization layers informed by acquisition metadata, lightweight domain adapters, adversarial domain-invariance losses, and self-normalizing modules (e.g., SeNMo-style architectures [54]) to maintain performance across institutions and scanners while avoiding spurious correlations. Finally, clinically constrained outputs and interpretability are increasingly built into the design of oncologic FMs. Rather than producing only generic embeddings; these models aim for representations and prediction heads that align with clinically interpretable entities, standardized response criteria, risk scores, staging categories, or biologically meaningful latent factors. As discussed in Section 5.1.2.A, this includes not only saliency and attention maps but also structured rationales and chain-of-thought–style explanations (of the kind popularized by modern LLMs such as Gemini, Claude, DeepSeek-R1, and the ChatGPT) that can, in the future, be grounded in specific image regions, laboratory values, or omics features. Together, these algorithmic directions illustrate that medical FMs require domain-optimized model design, not merely medical datasets, to be clinically useful and trustworthy.

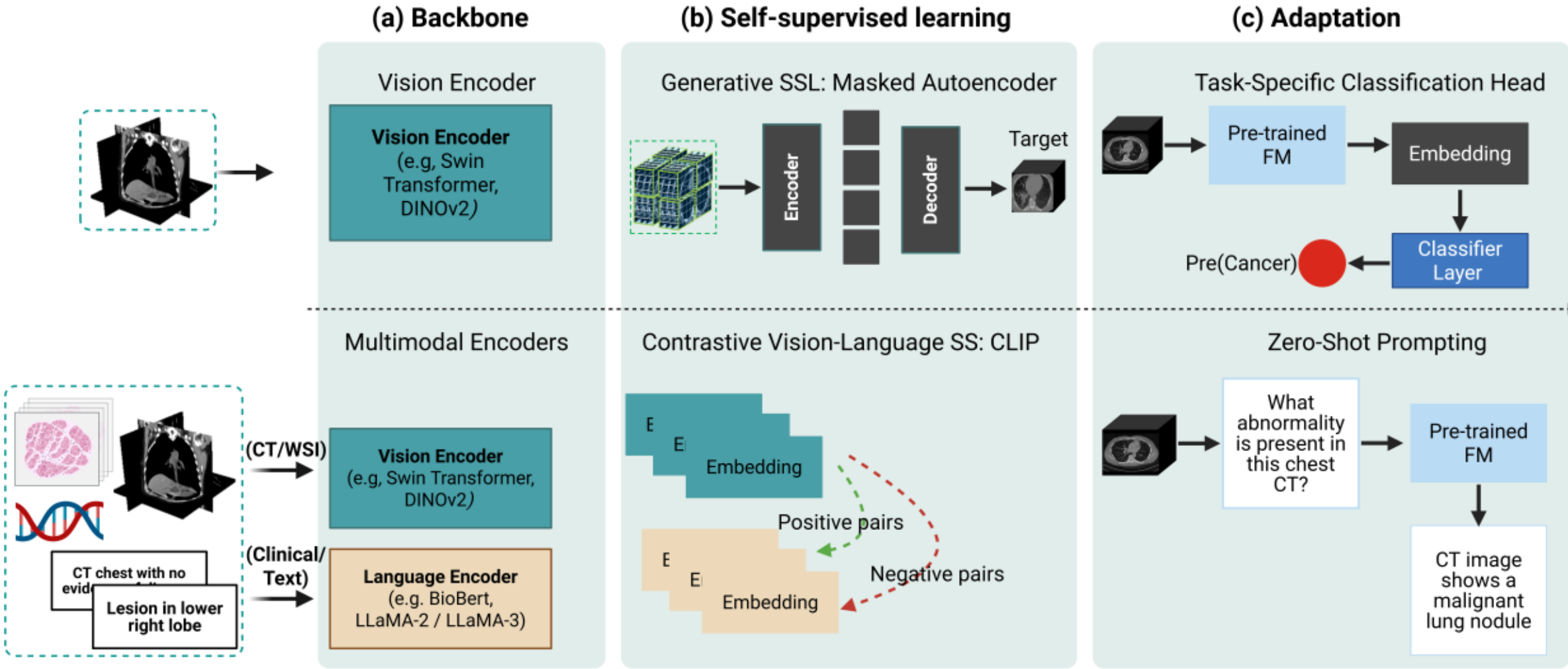

**Figure 7**: Algorithmic design of Cancer multimodal FMs. (a) Domain-specific backbones, including unimodal vision encoders and multimodal vision–language encoders, b) Self-supervised pretraining using generative masked-autoencoder objectives and contrastive vision–language learning, and (c) Downstream adaptation via task-specific heads or zero-shot prompting

### *5.5. Integrating Spatial Phenotypes with Molecular Data*

The convergence of two powerful technologies enables an unprecedented, spatially resolved view of cancer biology. The first, digital histopathology, transforms tissue slides into giga-pixel WSIs, providing a high-fidelity spatial canvas of cellular architecture and morphology—a gold standard for diagnostics [178, 179]. The second, spatial omics, overlays this canvas with precise molecular data. Technologies such as Visium HD [180], digital spatial profiling (DSP) [97, 181], and spatial molecular imaging (SMI) [182] now allow for the direct mapping of transcriptomic or proteomic profiles onto tissue structures, creating a one-to-one link between molecular state and histological context [183].

This technological synergy has catalyzed two principal paradigms for data integration. The first is in silico molecular inference, where DL models are trained to predict molecular features directly from standard H&E stained WSIs, effectively acting as a computational microscope to bridge histology and genomics when direct spatial data is unavailable. Seminal works like HE2RNA and HE2Gene, often leveraging advanced architectures with self-attention, have demonstrated the potential to infer gene expression profiles from tissue morphology alone [184-188]. The second, more direct paradigm is the fusion of WSIs with concurrently measured spatial omics data. This approach allows researchers to directly correlate molecular activity with histological features like tumor invasion and immune infiltration, offering a more mechanistic view of the tumor microenvironment [183, 189-191].

Both paradigms, however, grapple with significant technical hurdles. A primary challenge is the immense data heterogeneity—reconciling giga-pixel image data with sparse, high-dimensional omics matrices. To manage the scale of WSIs, Multiple-Instance Learning (MIL) has emerged as a standard strategy, dividing images into thousands of smaller patches (instances) whose information is then aggregated for joint analysis with omics data [192, 193]. A second major bottleneck is the availability of well-curated, paired (matched) datasets in which histology, spatial omics, and/or radiology are measured on the same tissue specimen or patient. In many practical scenarios, researchers instead rely on unmatched datasets—for example, bulk RNA-seq from one cohort and WSIs from another—where integration must proceed via statistical alignment, domain adaptation, or shared latent spaces rather than one-to-one correspondence. Such unmatched designs greatly expand the usable data pool but introduce additional assumptions and potential biases (e.g., differences in sampling, preparation protocols, or patient mix)[194], underscoring the particular value of genuinely matched, spatially resolved resources such as TCGA, the Human Tumor Atlas Network (HTAN), STimage-1K4M [195], and Hest-1K [196] for training and validating integrative models [183, 189-191, 197].

Finally, the principles of spatial-molecular fusion are now extending from the microscopic scale of pathology to the macroscopic view of radiology. FMs for volumetric imaging, such as those for CT, are pivotal in this expansion [155, 156]. By transforming massive 3D CT volumes into compact, information-rich embeddings, these models create an "omics-ready" representation of the tumor and its environment. This radically simplifies the technical challenge of integrating radiological phenotypes with genomics or proteomics, complementing the microscopic view provided by histopathology and accelerating the development of holistic biomarker models for precision medicine.

## 6. Open Challenges and Future Directions

### *6.1. Open Challenges.*

The integration of multimodal data, powered by advanced methods, holds immense promise for revolutionizing cancer research, diagnosis, prognosis, and personalized treatment. However, despite significant advancements, major obstacles remain. Crucially, many of these challenges only fully manifest when models are deployed in real-world clinical environments, where technical limitations intersect with regulatory constraints, workflow integration issues, and human factors. These challenges are systematically delineated in the following sections.

#### *6.1.1. Data-Centric Challenges*

The performance, reliability, and ultimate clinical utility of any AI model are inextricably bound to the fidelity, structure, and integrity of the data upon which it is trained. The most sophisticated algorithm cannot overcome the limitations of flawed or poorly harmonized inputs. In the context of multimodal data fusion, many challenges that manifest as algorithmic failures are, in fact, symptoms of deeper, unresolved data-level issues. These foundational gaps, from intrinsic data heterogeneity to the practicalities of missing data and computational scale, represent the primary bottleneck to progress in the field. They arise both within individual modalities and, critically, in their joint representation, where structurally divergent signals must be coherently aligned.

*A. The Conundrum of Data Heterogeneity and Standardization*

The central data-centric challenge is reconciling the deep heterogeneity of multimodal datasets. This is not simply concatenating feature vectors, but bridging fundamental differences in data structure, statistics, and semantics. Genomics is often discrete (e.g., somatic mutations), transcriptomics continuous (expression levels), and proteomics semi-quantitative; integrating these with imaging, where information is encoded in spatially correlated pixels/voxels, is far more complex than correcting batch effects within a single modality [198]. Naïve fusion of such disparate feature spaces risks substantial information loss.

Compounding this intrinsic heterogeneity is a lack of standardization in acquisition, processing, and annotation. Across and within institutions, changes in protocols, sequencing platforms, imaging scanners, and bioinformatics pipelines introduce technical artifacts and batch effects that powerful but naïve models may misinterpret as biology [199, 200]. This "standardization void" limits the usefulness of public datasets and undermines reproducibility and generalizability: models trained on one pipeline often degrade sharply on another, eroding clinician trust, complicating regulatory approval, and contributing to a potential reproducibility crisis [201].

These problems are especially acute for imaging-derived features, where radiomics and quantitative imaging have shown strong shifts with vendor, reconstruction kernel, slice thickness, contrast phase, staining protocol, and digitizer settings. Large multimodal FMs pooling CT, MRI, PET, and WSI thus face not only biological variability but also pronounced acquisition- and device-induced shifts; if unmitigated, they may encode scanner identity rather than tumor biology. Addressing this requires a two-pronged strategy that couples' data-centric standardization with FM design. On the data side, widely adoptable primitives can reduce trivial domain artifacts: systematic capture of scanner and protocol metadata, physical or digital phantoms for cross-site calibration, intensity and stain normalization, and small benchmark calibration sets spanning vendors and institutions [176]. On the model side, cancer-specific, domain-aware FMs are needed, using shared backbones with lightweight domain adapters, conditional normalization layers parameterized by acquisition metadata, adversarial or contrastive objectives that encourage domain-invariant representations, and self-normalizing modules (e.g., SeNMo-style architectures [54]) that explicitly regularize across cohorts. Federated learning and multi-site pretraining further allow institutions to contribute data without centralization while still shaping a common representation space [136].

*B. The Curse of Dimensionality and the Signal-to-Noise Crisis*

The integration of multiple omics layers dramatically exacerbates the classic "curse of dimensionality," or the $p > n$ problem, where the number of features ($p$) vastly exceeds the number of samples or patients ($n$). While transcriptomic data alone may comprise tens of thousands of features, combining it with proteomics, genomics, and metabolomics can easily push the feature space into hundreds of thousands or millions. In this regime, models are highly prone to overfitting—capturing spurious cohort-specific correlations rather than generalizable biology—unless strong dimensionality reduction and feature selection are applied.

The challenge, however, is not only dimensionality but also cross-modal amplification of noise. Each modality introduces its own error profile (e.g., sequencing artifacts, antibody cross-reactivity, sample degradation, imaging noise), and their superposition can create complex, non-linear confounders that simple, modality-specific denoising cannot remove. This "signal-to-noise crisis" calls into question the default assumption that "more modalities are always better." Beyond a certain point, adding a noisy omics layer may inject more confounding variance than useful signal, leading to negative returns on integration. A key future direction is therefore to integrate selectively rather than exhaustively, using meta-learning or data-valuation schemes that estimate the marginal signal-to-noise contribution and cross-modal complementarity of each modality for a given task—identifying, for example, when imaging, spatial transcriptomics, and routine clinical variables combine synergistically and when specific modalities should be downweighted or excluded to preserve robustness [115].

*C. Missing Modalities*

In benchmark datasets, complete multimodal profiles per sample are often assumed; in real clinical settings this is rare. Patients frequently lack specific molecular tests or imaging scans due to cost, contraindications, sample availability, or logistics [107], making partially observed data a pervasive barrier to integration [94]. Training models that remain reliable under incomplete data is therefore a core requirement for any clinically viable AI system, not a special case.

A variety of computational strategies have been developed to address this challenge. Probabilistic frameworks, such as Multi-Modal Factor Analysis (MOFA) [202] and total Variational Inference (totalVI) [203], employ Bayesian latent variable models to infer a shared set of underlying biological factors from the available data, gracefully handling instances where entire modalities are missing for certain samples. In the DL domain, techniques like multimodal dropout deliberately omit random modalities during the training process, forcing the model to learn redundant and resilient representations, thereby enhancing its ability to make predictions when faced with incomplete data at inference time [204]. Generative models (VAEs and GANs) can impute missing views by learning non-linear cross-modal relationships from complete cases [205]. For the unique challenges of sparse and incomplete single-cell data, specialized methods like Joint Sparse Non-negative Matrix Factorization (JSNMF) have been designed to integrate transcriptomic and epigenomic profiles effectively [206, 207], while methods such as WCluster [208, 209] and pipelines like Panpipes [210] demonstrate how network-based clustering and modular workflows can support subtype discovery and large-scale analysis despite incomplete modalities [211].

However, most existing approaches implicitly assume data are "missing at random". In clinical practice, missingness is often informative: tumour genomics may be absent because a patient is too frail for biopsy, or PET imaging may be missing because care is delivered in a lower-resource setting. In such cases, missingness correlates with health status, socioeconomic context, and patterns of care [194]. Naïve imputation can therefore discard predictive signal and introduce systematic bias (e.g., underestimating risk in frail patients by filling in genomics from healthier cohorts). A key frontier is the development of missingness-aware models that treat the data-availability matrix itself as input, learning from which modalities co-occur or are absent in specific subgroups and how these patterns reshape the complementary information landscape. This shifts the focus from pure imputation to holistic patient modelling, where the process of data acquisition carries as much context as the measurements themselves.

Beyond clinical and logistical factors, patterns of missing data are also shaped by regulatory and technical constraints in multi-institutional cohorts. Privacy and data-protection frameworks such as Health Insurance Portability and Accountability Act (HIPAA) in the United States and the General Data Protection Regulation (GDPR) in Europe frequently require aggressive de-identification (e.g., removal of dates or geographic identifiers) or outright prohibition of centralizing raw imaging, omics, and EHR data [212]. To comply, institutions increasingly resort to federated or distributed-learning setups in which data remain local and only model updates are shared, creating a form of "systematic missingness" from the perspective of a global model that never sees the full variable space [136]. Data-localization policies, heterogeneous consent models, and institutional review board constraints further fragment the landscape, leading to multimodal cohorts that are "artificially incomplete" even when all modalities exist somewhere in the network [213]. Addressing this regulatory layer of missingness will require not only methodological advances such as federated and privacy-preserving learning, secure linkage of distributed datasets, and robust modeling under partial observability, but also harmonized governance frameworks that enable responsible data sharing without compromising patient privacy [136].

*D. The Scalability*

The sheer scale of modern biomedical data is pushing the limits of conventional computing infrastructure. A single high-resolution whole-slide pathology image can approach a terabyte in size, and integrating such data with multiple omics layers for large patient cohorts generates datasets of peta-scale proportions. The algorithms required to analyze this data often involving high-dimensional statistics, graph analytics, and DL are computationally intensive. Consequently, High-Performance Computing (HPC) has transitioned from a specialized tool to an essential backbone for cutting-edge multi-omics/multimodal research. HPC architectures, which leverage parallel processors, massive, shared memory, and high-speed interconnects, can reduce analysis times from weeks or days to mere hours, making computationally demanding and more accurate integration algorithms feasible for the first time.

However, the reliance on HPC introduces its own set of formidable challenges. First, technical bottlenecks related to data movement, memory limitations, and interoperability between different software frameworks continue to impede the development of seamless and efficient analysis pipelines. Second, and more critically, the high cost and specialized expertise required to build and maintain HPC infrastructure creates a significant barrier to entry. These risks create an "HPC access gap," where cutting-edge research becomes concentrated in a few elites, well-funded academic centers, thereby stifling innovation at smaller institutions and widening global healthcare and research disparities. Third, the immense power of HPC enables the training of ever larger and more complex "black box" models, which further exacerbates the critical challenge of model interpretability.

This increasing dependence on HPC is fundamentally reshaping the culture of biomedical research. It necessitates a new, deeply integrated tripartite collaborative model, bringing together domain experts (biologists and clinicians), data scientists (ML and statistics experts), and computational scientists (HPC and systems experts). This shift away from the traditional single-investigator lab model is a significant organizational and cultural evolution. It also raises a profound strategic question for the future of the field: should the primary focus be on building ever larger and more powerful supercomputers, or on designing more computationally efficient, "greener" algorithms that can deliver state-of-the-art results on more modest and accessible hardware? Navigating this trade-off between hardware scale and algorithmic intelligence will be critical for ensuring the democratization and long-term sustainability of computational medicine. For multimodal FMs that jointly encode imaging, omics, and clinical data, this trade-off is particularly acute: models that fully exploit cross-modal interactions tend to be among the most computationally demanding, which risks restricting their development, evaluation, and deployment to a small number of well-resourced centers. As illustrated in **Figure 8**, representative models across modalities show an approximately exponential growth in parameter count over the last decade, and the bubble sizes provide a relative proxy for per-inference compute and energy requirements based on published FLOP counts and models.

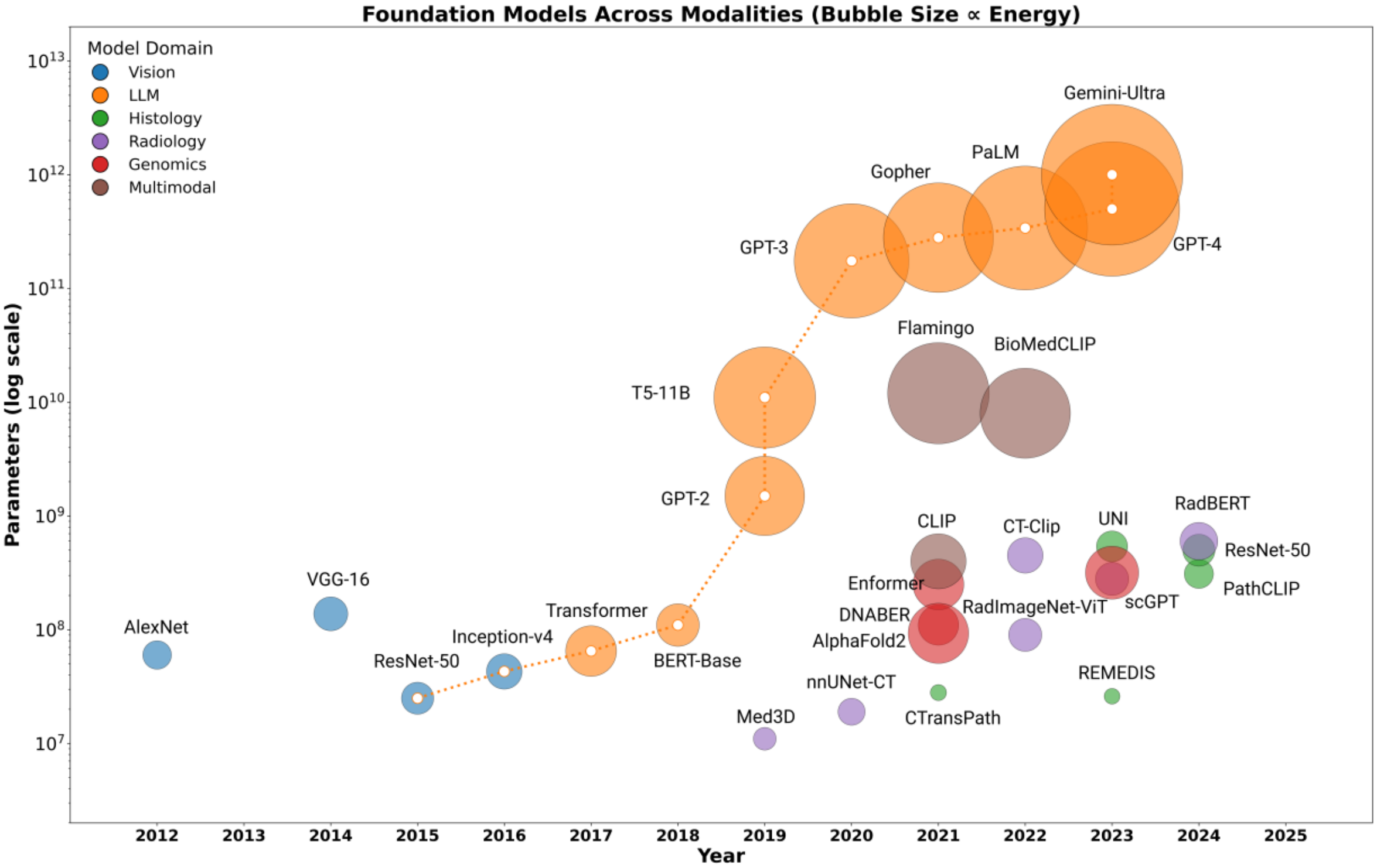

**Figure 8:** Growth of foundation models across modalities (bubble size ∝ relative compute/energy proxy). Parameter values and release years are taken from original model papers and technical reports. Bubble area is proportional to an order-of-magnitude proxy for per-inference energy cost, computed from reported or commonly cited FLOP estimates under a fixed hardware efficiency assumption. The dotted line summarizes the approximate exponential increase in parameter count over time; bubble sizes are intended as qualitative, not exact, energy comparisons.

Considered as a whole, these data-centric challenges underscore that multimodal integration is not simply the union of several independent analyses. Instead, it requires carefully engineered data ecosystems in which heterogeneous modalities are collected, standardized, and curated with integration in mind from the outset, so that their complementary strengths, molecular resolution, spatial context, and longitudinal clinical information, can be jointly leveraged rather than competing or confounding one another.

From this perspective, current multimodal FMs do not automatically resolve these data-centric difficulties and, in some settings, may exacerbate them. Large, flexible architecture are particularly prone to encoding site-, scanner-, or pipeline-specific artifacts when training data are only partially harmonized. In practice, a smaller, well-curated unimodal or bi-modal model can be more reliable than a nominally "multimodal" FM trained on heterogeneous, weakly standardized inputs. Thus, the advantages of FMs over classical pipelines are contingent on rigorous data curation and harmonization and cannot be assumed to hold under arbitrary real-world data conditions.

*E. Data Accessibility, Governance, and Regulatory Constraints*

Even if technical challenges around heterogeneity and standardization are solved, the development of clinically meaningful cancer FMs is fundamentally constrained by data accessibility and governance. Public, open-source datasets remain indispensable for method development and benchmarking, but they capture only a narrow slice of real-world oncologic practice, often biased toward tertiary centers, specific imaging vendors [194, 214]. Truly robust FMs will require access to large-scale, longitudinal, and demographically diverse clinical cohorts, including raw imaging, detailed treatment histories, and high-resolution molecular profiles. Frameworks such as HIPAA, GDPR, and analogous national regulations constrain secondary use, cross-border transfer, and cross-modal linkage; within a single health system, radiology, pathology, genomics, and clinical records may even sit under different consent and governance regimes [212, 213].

If pretraining remains limited to a few permissive centers or regions, cancer FMs will systematically under-represent many populations, tumor types, and health systems, undermining both equity and generalizability. Overcoming this requires treating governance as a design constraint, not an afterthought: technical strategies (federated and secure multi-party learning, differential privacy [136]) must be coupled with harmonized data-use agreements and consent frameworks. Only this combination can enable multi-institutional, multimodal training at scale while remaining compatible with legal, ethical, and societal expectations.

6.1.2. Model-Centric Challenges

Even with access to perfectly harmonized, complete, and high-quality data, the intrinsic characteristics and operational complexities of advanced DL and FMs models present their own set of significant challenges. The focus here shifts from the data to the algorithms themselves. The very properties that make models like deep neural networks so powerful, their ability to learn complex, non-linear relationships in high-dimensional space, also make them opaque, difficult to fuse with disparate data types, and prone to failures in generalization. Bridging the gap between algorithmic capability and clinical reality therefore requires models that are not only accurate, but also interpretable, robust, and operationally manageable. For multimodal FMs in oncology, this entails designing architectures and training schemes that can explicitly model cross-modal interactions (e.g., through cross-attention, hierarchical fusion, or shared latent spaces [90-92]), handle missing or asynchronous modalities, and propagate calibrated uncertainty across all input channels in a way that is compatible with clinical decision-making.

Despite the enthusiasm surrounding multimodal FMs, empirical evidence for their superiority over task-specific DL or classical ML approaches in oncology remains limited. Most published evaluations are retrospective, often single center or few centers, and rely on internal or closely related external test sets. Head-to-head comparisons against strong, carefully tuned conventional baselines are still relatively rare, and truly prospective studies are even scarcer. Consequently, claims of performance gains should be interpreted as promising but preliminary, rather than definitive proof that FMs consistently outperform existing methods across tasks, institutions, and patient populations.

*A. Model Interpretability and Explainability*

The opacity of many state-of-the-art AI models, specifically transformers or FMs, is arguably the single greatest barrier to their widespread adoption in clinical practice. In high-stake medical decision-making, a prediction, no matter how accurate it is, is insufficient. Clinicians, regulators, and patients require a comprehensible and verifiable reasoning process to establish trust, ensure accountability, and facilitate genuine scientific discovery [215]. This "black box" problem creates a profound trust deficit, as clinicians are justifiably hesitant to base critical patient care decisions on algorithmic outputs they cannot scrutinize or understand.

To address this, explainable AI (XAI) has emerged [216], but it is crucial to understand the nuanced distinction between two key concepts. Interpretability refers to models that are inherently transparent by design. These models, such as linear regression, decision trees, or rule-based systems, provide a clear relationship between inputs and outputs that is directly understandable by a human user without supplementary tools [217]. Explainability, in contrast, typically refers to the application of post-hoc techniques to an already-trained, opaque model to provide an approximation or justification of its behavior. These methods can be model-specific, like Gradient-weighted Class Activation Mapping (Grad-CAM) for convolutional neural networks [218], or model-agnostic, like Local Interpretable Model-agnostic Explanations (LIME) [219] and SHapley Additive exPlanations (SHAP) [220], which probe the model's input-output behavior to assign importance scores to features.

While post-hoc explainability methods are a necessary step, they are an imperfect patch rather than a complete solution. These explanations can be fragile, potentially misleading, and may not faithfully reflect the model's true internal logic. They explain what features the model used but not necessarily why in a way that aligns with human causal reasoning [221]. A complementary line of work, particularly in large language and multimodal models (e.g., Gemini, OpenAI o-series, Claude, DeepSeek-R1), is the use of self-generated reasoning traces or "chain-of-thought" explanations [222]. These systems output an explicit, stepwise textual rationale alongside the prediction, bringing model behavior closer to how clinicians articulate reasoning and potentially aid error checking and documentation [221]. However, such traces are not guaranteed to be faithful—they may serve as post-hoc rationalizations or even hallucinate intermediate steps [223]. For safety-critical applications, including multimodal FMs in oncology, it will be essential to rigorously evaluate the fidelity of these reasoning traces and to anchor each step in verifiable evidence from the input (e.g., specific image regions, laboratory values, or omics features) before relying on them in clinical workflows [215].

These challenges are particularly acute for multimodal cancer FMs, whose very strength lies in combining heterogeneous inputs within deep, highly parameterized architectures. Cross-modal attention shared latent spaces, and hierarchical fusion layers further obscure the mapping from raw inputs to predictions, making it difficult to attribute specific decisions to interpretable biological or clinical factors [115]. At present, there is little systematic evidence that existing XAI techniques or reasoning-trace mechanisms provide faithful, regulator-ready explanations for such systems in oncology [220, 222]. As a result, interpretability remains a major unresolved barrier that tempers any claims about the immediate clinical readiness of multimodal FMs.

This realization is driving a paradigm shift away from simply "explaining AI" towards creating "collaborative AI". This involves a deeper integration of clinical expertise throughout the entire model's lifecycle, a concept known as the "human-in-the-loop" approach. Instead of clinicians being passive recipients of prediction and static explanation, they become active participants in a collaborative reasoning process [215]. The future of clinical AI may lie in systems that do not just provide an answer, but engage in a dialogue, presenting a differential diagnosis, the evidence supporting each possibility, a measure of confidence, and crucially, highlighting instances where its prediction is based on sparse or conflicting data. Such systems reposition AI from an opaque oracle to a cognitive partner that shares part of the reasoning workload while leaving value-laden decisions with clinicians.

*B. Robustness and Generalizability*

A model that achieves stellar performance on a curated benchmark dataset may be celebrated in a research paper, but its true value is only realized if it performs reliably in the chaotic and heterogeneous environment of real-world clinical practice. A model's accuracy on a held-out test set from the same data distribution is a poor proxy for its clinical utility. The most significant challenge in translating AI models from the lab to the clinic is the "generalizability gap", the sharp drop in performance when a model is deployed on data from new hospitals, different imaging scanners, or diverse patient populations not represented in the original training data.

This failure to generalize often stems from models learning "data leakage" by exploiting spurious correlations or technical artifacts present in the training data. For example, a model might learn to associate a subtle image artifact unique to a particular MRI scanner with a specific cancer subtype, achieving near-perfect classification on the training data for entirely the wrong reasons. When deployed at a new hospital with a different scanner, the model fails completely. This emphasizes the critical importance of both rigorous data harmonization (*as commented in Section 6.1.1*) and, crucially, robust validation protocols. Standard k-fold cross-validation is insufficient. True validation requires testing on large, independent, external datasets from multiple institutions, a practice that is becoming the gold standard for assessing real-world performance.

In the end, generalizability cannot be solved by data collection alone; it is computationally and logistically infeasible to collect training data from every possible clinical setting. The solution must also be algorithmic. This points to the vital importance of research into ML techniques for "domain adaptation" and "domain generalization." These methods aim to build models that are inherently robust to "domain shift"[224]; the change in data distribution between training (source domain) and deployment (target domain). These approaches explicitly train a model to be invariant to domain-specific features (e.g., scanner-specific image textures) while focusing only on the domain-invariant features that represent the true underlying biology. This can be achieved through techniques like adversarial training, where a sub-network of the model attempts to predict the data's domain of origin (e.g., which hospital it came from), while the main predictive model is trained to generate representations that make this task impossible. By forcing the model to discard domain-specific information, it learns a more robust and generalizable representation of the disease. In multimodal FMs, these ideas extend naturally to learning representations that are both cross-site and cross-modal invariant, so that imaging, omics, and clinical text contribute to a stable shared disease representation. The future of reliable clinical AI lies not just in amassing "big data," but in developing "domain-aware" learning architectures that are built from the ground up to anticipate and master the heterogeneity of the real world.

Notably, increasing model scale does not automatically close this generalizability gap. Large multimodal FMs can still be overfit to subtle, site-specific artifacts or demographic imbalances, and may in fact be more capable of memorizing idiosyncratic patterns in limited datasets. Current reports of improved cross-site robustness are encouraging but remain based largely on retrospective benchmarks with constrained diversity. There is, as yet, insufficient prospective, multi-institutional evidence to conclude that multimodal FMs are inherently more robust or transportable than well-regularized task-specific models, particularly when deployed in resource-limited or data-sparse settings.

*C. Accountability*

The demand for transparent, collaborative AI directly informs the critical issue of accountability. When an AI system contributes to a clinical decision, establishing clear lines of responsibility is not merely a legal or administrative formality; it is a fundamental requirement for ethical practice and patient safety [222]. Accountability frameworks must address the entire lifecycle of the AI model, from the initial data curation and potential biases embedded within it, to the validation process and post-deployment surveillance {Aldea, 2025 #65}.

In multimodal FMs context, this challenge is magnified. The complexity of the model and the vastness of the data it integrates can diffuse responsibility, making it difficult to pinpoint the source of an erroneous or harmful recommendation. Is the onus on the model developers, the institution that deploys the system, or the clinician who ultimately accepts its guidance? Therefore, establishing robust governance structures is paramount. These structures must define roles, responsibilities, and accountability, ensuring that despite the complexity of the underlying technology, a clear path to redress and continuous improvement exists. Without such a framework, even the most interpretable and collaborative systems will fail to gain the institutional and societal trust necessary for their integration into routine cancer care, leaving their potential to enhance clinical decision-making unrealized.

*D. Algorithmic Bias, Fairness, and Health Equity*

Algorithmic bias and fairness are not abstract concerns but measurable properties of model behavior under distributional shifts [194, 214]. In multimodal oncology FMs, several sources of bias frequently coexist: sampling bias (over-representation of patients from high-income regions or tertiary centers), measurement bias (systematic differences in scanners, staining protocols, or sequencing platforms), label bias (subjective or inconsistently applied clinical labels), and representation bias (under-representation of specific demographic or clinical subgroups in the feature space) [194]. Multimodal integration introduces an additional, less discussed dimension: modality-access bias. Advanced imaging (e.g., PET/CT, whole-slide imaging) and multi-omics profiling are more commonly available in well-resourced settings, such that the "full" multimodal stack is inherently skewed toward already advantaged patients.

These biases can lead to heterogeneous error profiles across subgroups, even when global metrics appear satisfactory [217]. For instance, a survival FM trained on imaging–omics–clinical data from a comprehensive cancer center may achieve strong overall C-index yet substantially underperform in patients from community hospitals where molecular profiling is sparse and imaging protocols differ. In a multimodal setting, the interaction between modalities can also amplify bias: if certain combinations of modalities (e.g., WSI + RNA-seq + detailed clinical text) are predominantly observed in younger, insured, or majority-ethnicity patients, the learned latent space may devote the highest capacity to these strata, while systematically degrading performance on under-represented groups or reduced-modality pathways.

From a technical standpoint, fairness assessment in multimodal FMs should therefore go beyond a single aggregate AUC or C-index. At minimum, models should be evaluated using subgroup-stratified metrics (e.g., AUC, calibration error, and Brier score stratified by sex, age bands, race/ethnicity where available, site, and stage), and, where appropriate, formal fairness criteria such as equal opportunity (similar sensitivity across groups for positive cases), equalized odds (similar sensitivity and specificity), or worst-group risk (distributionally robust evaluation). In multimodal contexts, it is also informative to report performance as a function of modality availability patterns (e.g., imaging-only vs. imaging+omics vs. omics+clinical) to detect systematic failures when certain modalities are missing or degraded—an issue tightly coupled to real-world inequities in diagnostic access. Mitigation strategies likewise need to be adapted to the multimodal setting. At the data level, approaches such as reweighting, stratified sampling, or targeted enrichment of under-represented subgroups can reduce imbalance, while care must be taken not to overfit to small sub-cohorts (**Figure 9**). At the model level, fairness-aware objectives (e.g., adding penalty terms for subgroup performance gaps), adversarial debiasing (training latent representations that obfuscate protected attributes while preserving predictive signal), and distributionally robust optimization (explicitly optimizing performance under worst-case subpopulation shifts) provide principled ways to trade off average accuracy against equity. Importantly, these techniques should be evaluated transparently, reporting not only gains in fairness but also any associated changes in overall performance and uncertainty [177].

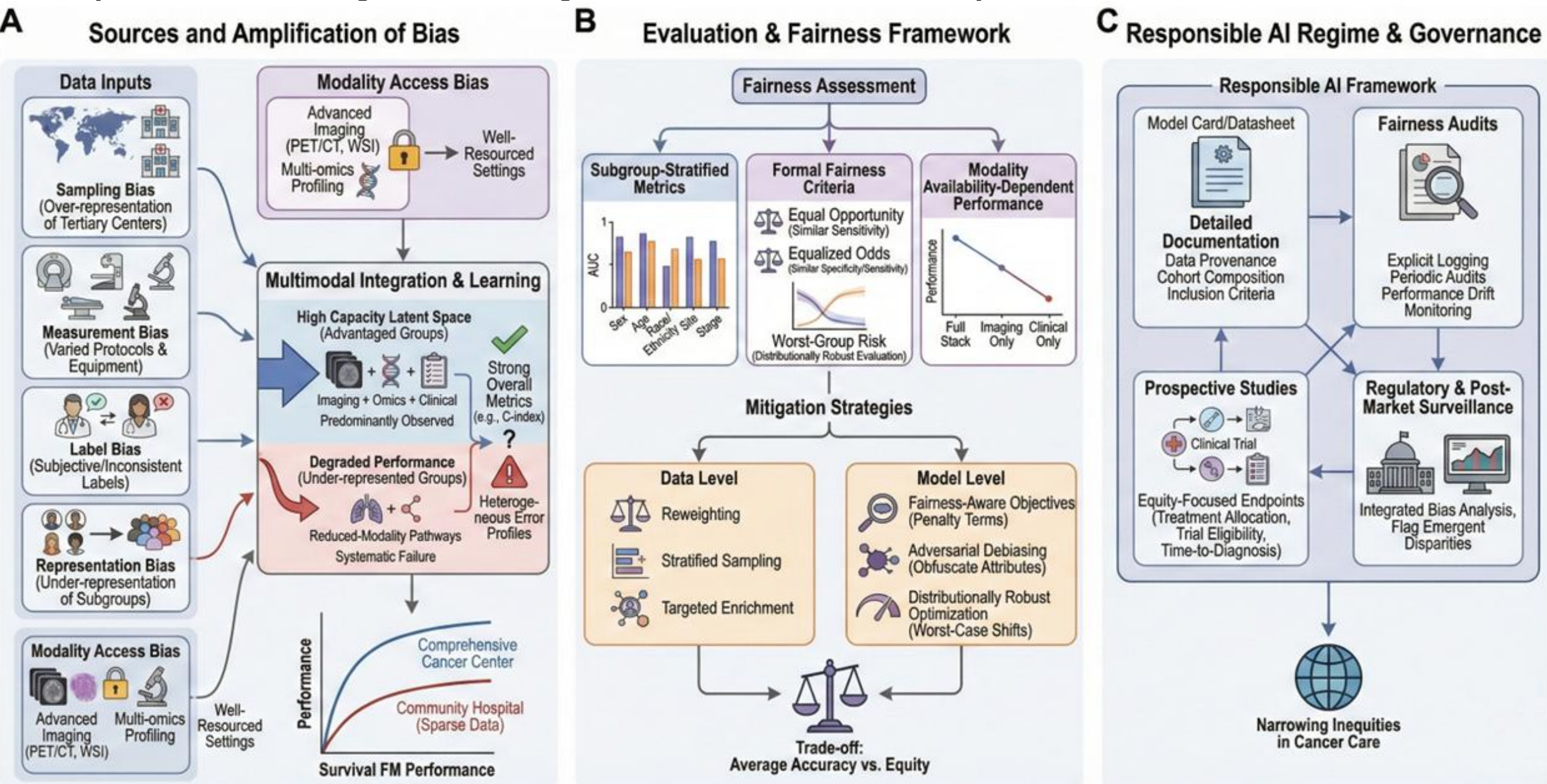

**Figure 9**: Algorithmic Bias, Fairness, and Health Equity in Multimodal Oncology FMs

Embedding multimodal FMs within a responsible AI framework requires operational components in addition to algorithms. Detailed documentation of data provenance, cohort composition, and inclusion/exclusion criteria (e.g., through model cards and dataset “datasheets”), explicit logging of how each modality contributes to the decision, and governance structures for periodic fairness audits are all essential [194]. Prospective studies should incorporate equity-focused endpoints such as differential impact on treatment allocation, trial eligibility, or time-to-diagnosis across subgroups, rather than treating fairness solely as an offline metric. Finally, bias and fairness analyses should be integrated into regulatory review and post-market surveillance, with monitoring pipelines that flag performance drift or emergent disparities as practice patterns, populations, and modality availability evolve. Only under such a responsible AI regime can multimodal FMs realistically contribute to narrowing, rather than widening, existing inequities in cancer care.

*E. Fusing Imaging with Molecular Data*

The fusion of spatial data from medical imaging with non-spatial data from omics represents a unique and formidable frontier in multimodal integration. The challenge is not merely computational but deeply conceptual, requiring the invention of novel frameworks that can bridge the vast semantic and structural gap between pixels and molecules. A primary technical hurdle is the co-registration and alignment of data acquired at vastly different scales and resolutions. For example, accurately mapping a 50-micron spot from a ST array to the precise morphological features within a giga-pixel whole-slide histopathology image is a non-trivial task where even minor misalignments can lead to profoundly incorrect biological conclusions [24].

This challenge is exemplified by the integration of ST with histopathology. While ST provides unprecedented insight into the spatial organization of gene expression within a tissue, its utility is hampered by persistent technical issues, most notably the accuracy of cell segmentation. Inaccuracies in delineating cell boundaries can create artifacts that make cells appear more transcriptionally similar to their immediate neighbors than they truly are, potentially obscuring genuine cell-cell interaction signals and confounding biological interpretation. The goal of this integration, known as "pathogenomics", is similar to radiogenomics, which develop models that can predict molecular states, such as the presence of a specific gene mutation or the activity of a signaling pathway, directly from routine, non-invasive clinical images. Achieving this requires AI architectures, such as multi-branch neural networks, capable of learning a shared latent representation where features from these incongruent modalities can be meaningfully compared and fused. In practice, this argues for models that not only align representations across modalities, but also explicitly model cross-modal attention, enforce consistency between image-derived and omics-derived predictions, and surface discordances that may signal biological novelty or technical failure. **Figure 10** illustrates how spatial imaging and omics data can be jointly processed through a unified multimodal model to yield clinically actionable pathogenomic insights.

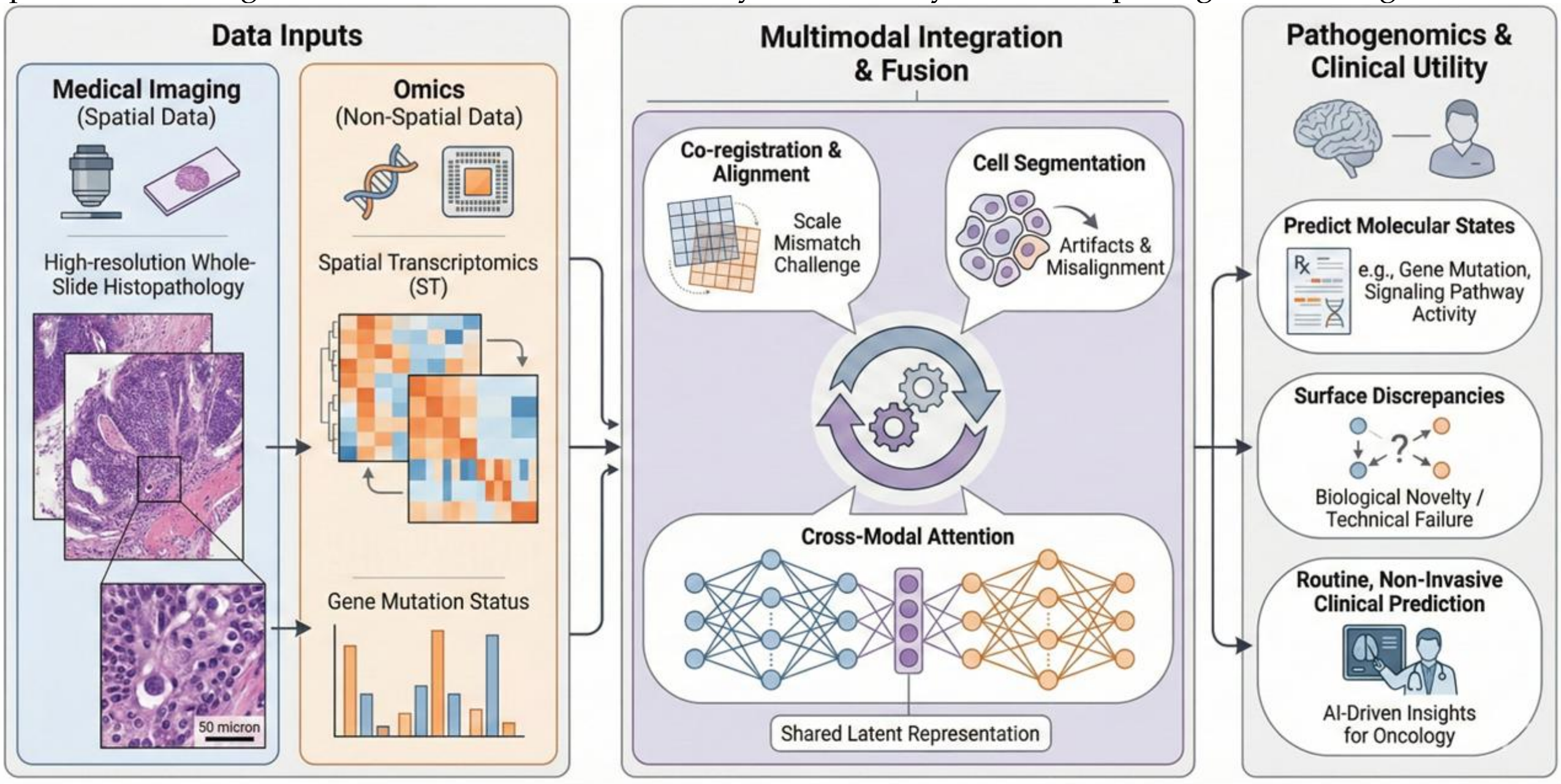

**Figure 10**: Schematic overview of a multimodal pathogenomics pipeline.

*F. Clinical Workflow Integration, Regulation, and Real-World Deployment*

Even the most sophisticated multimodal FMs [56, 149-156] remains purely academic until it can be safely, reliably, and efficiently embedded within routine clinical workflows. Practical implementation poses a distinct set of challenges that go beyond algorithms and data [225]. At the workflow level, AI systems must integrate with existing hospital infrastructure (e.g., EHR, PACS, laboratory information systems) without adding unacceptable latency or cognitive burden to already overextended clinicians. Multimodal models that require simultaneous access to imaging, pathology, and multi-omics data must respect the asynchronous and often fragmented nature of real-world clinical pathways, where different modalities become available at different time points and sometimes not at all.

From a regulatory perspective, multimodal FMs deployed for diagnosis, prognosis, or treatment recommendation are typically classified as software as a medical device (SaMD). This entails rigorous documentation of training data provenance, performance characteristics across subgroups, model update policies, and post-market surveillance procedures. For systems that continuously learn from new data or adapt to local practice patterns, regulators increasingly require explicit mechanisms for change control and re-validation, as well as clearly defined processes for monitoring model drift and clinical impact over time. Experience with the first wave of FDA-cleared AI/ML-based medical devices has already demonstrated that regulatory approval is a necessary but far from sufficient condition for sustained clinical value: tools that perform impressively under carefully controlled conditions may show attenuated benefit, performance drift, or inequitable error patterns when exposed to the full heterogeneity of real-world practice [201]. These lessons are specifically salient for multimodal FMs, whose complexity and dependence on multiple data streams amplify both the potential for impact and the risks associated with unmonitored deployment. Beyond regulatory clearance, successful translation also depends on alignment with reimbursement pathways (e.g., procedure codes, value-based care metrics) and clear medico-legal guidance on how algorithmic recommendations should be incorporated into clinical responsibility frameworks.

Finally, successful deployment demands attention to human factors and organizational readiness [225]. In practice, many clinicians remain reluctant to rely on complex AI or FM-based systems, which makes it essential that they understand not only what the model predicts, but also under which conditions its recommendations are trustworthy or should be actively discounted, specifically for multimodal FMs whose outputs derive from intricate, non-intuitive interactions between imaging, molecular, and clinical signals. Implementation of science studies, prospective trials, and user-centered design are therefore essential complements to technical development and to the early proof-of-concept studies that currently dominate multimodal FM literature. Without such efforts, many multimodal FMs are likely to remain confined to isolated pilot projects and retrospective demonstrations rather than achieving the scalable, equitable impact on cancer care that they promise. Prospective impact evaluations, pragmatic trials, and real-world evidence studies that embed multimodal FMs into decision-support tools and quantify changes in survival, toxicity burden, time-to-diagnosis, clinician workload, and workflow efficiency will be key to closing the gap between technical performance and genuine clinical benefit.

For multimodal cancer FMs, this evidentiary gap is currently substantial. Most published models report technical metrics on retrospective cohorts, but few studies have demonstrated improvements in hard clinical endpoints, within prospective or pragmatic trial settings. Regulatory approvals specifically targeting multimodal FMs in oncology are essentially absent, and there is limited data on cost-effectiveness, workflow impact, or medico-legal acceptability at scale. Accordingly, the present state of the field should be viewed as an early translational phase: FMs are powerful research tools with clear potential, but their clinical utility and safety remain to be established through rigorous, longitudinal evaluation.

### *6.2. Future Directions*

While the field has made remarkable progress, the next phase of multimodal data integration in oncology will be driven by four priorities: (i) developing unified and dynamic foundation models, (ii) achieving mechanistically grounded interpretability, (iii) addressing data scarcity through principled generative modeling, (iv) implementing continual learning paradigms that keep models aligned with evolving clinical practice, and (v) reconciling global generalization with local, site-specific adaptation in clinical deployment. Below, we outline concrete directions that are technically feasible in the short-to-medium term and necessary for long-term clinical impact. The priorities outlined above are therefore aspirational and conditional: their eventual impact will depend on whether the community can address the data, model, and governance challenges, and on whether future studies demonstrate clear, reproducible benefits over established analytic pipelines in real-world settings.

*A. Beyond Specialization: Towards Unified and Dynamic Foundation Models*

Current medical imaging and multimodal FMs including UNI [149], PLUTO [150], GPFM [151], RudolfV [170], CTransPath [152], PRISM [153], CONCH [154], REMEDIS [57], TITAN [168], TMO-Net [107], P3GPT [147], Path-GPTOmic [148], MolFM [43], Nicheformer [143], SAM [56], CT-CLIP [155] and CT-foundation google [156] demonstrate substantial potential but remain largely specialized, typically constrained to a single modality or a narrow set of predefined combinations. A key next step is the development of unified multimodal FMs that can jointly encode imaging, multi-omics, and clinical data within a shared representational space. In the near term (3–5 years), actionable directions include:

- Designing cross-modal transformer architectures that support plug-and-play encoders for imaging, omics, and clinical text, with cross-attention layers explicitly modeling interactions between modalities.
- Establishing public benchmarks and challenge datasets where imaging is systematically linked to multi-omics and longitudinal clinical outcomes, enabling reproducible comparison of multimodal FMs.

In the longer term, models should move beyond static, single-timepoint representations toward spatio-temporal FMs that capture disease evolution:

- Spatially aware pathway modeling, where enriched molecular pathways are mapped onto tumor habitats, immune niches, or invasive fronts on histopathology or radiology.
- Temporally aware dynamic modeling, integrating serial imaging, treatment timelines, and temporal omics to predict resistance, relapse, and immune responses, potentially via temporal graph networks or dynamic Bayesian models will be essential to proactively inform treatment strategies rather than retroactively explaining failure.

*B. The Imperative for Mechanistic Interpretability*

For multimodal FMs to transition from promising prototypes to dependable clinical tools, interpretability must evolve from post-hoc correlation maps to mechanistically informative explanations. A concrete avenue is to use spatial omics and single-cell data as biological "ground truth" to verify whether image regions or latent factors prioritized by the model are enriched for specific cell states, pathways, or microenvironmental niches. Our recommendations for further research include:

- Developing joint visualization frameworks that align model attention maps with spatial omics readouts, allowing clinicians and biologists to interrogate whether predicted signatures are biologically plausible.
- Designing interpretable embedding spaces in which axes and clusters are explicitly annotated with known pathways, immune phenotypes, or stromal states, enabling mechanistic hypotheses to be generated rather than just black-box risk scores.
- Conducting prospective, clinician-in-the-loop studies that evaluate how multimodal explanations influence diagnostic confidence and treatment decisions, thereby quantifying the clinical utility of interpretability rather than treating it as a purely technical metric.

These efforts are immediately feasible with existing datasets and will be essential for regulatory acceptance and clinician trust.

*C. Generative Models as a Solution to Data Scarcity*

The scarcity of large, harmonized multimodal datasets remains a major bottleneck. Advanced generative models, progressing from early GANs to guided diffusion and flow-matching models, offer a principled route to augmenting and stress-testing multimodal pipelines, especially for rare cancer subtypes and under-represented populations. The near-term research directions include:

- Fine-tuning large, pretrained generative FMs to synthesize paired imaging–omics–clinical samples under explicit biological constraints (e.g., preserving known pathway structure or mutational patterns).

- Developing task-specific synthetic data protocols, where generative models are used not to replace real data but to: (i) rebalance class distributions, (ii) simulate edge cases (e.g., rare mutations), and (iii) probe model robustness to control distribution shifts.
- Establishing standardized evaluation frameworks for synthetic data that go beyond low-level similarity metrics, focusing instead on functional validation. For example, demonstrating that models trained with synthetic augmentation generalize better on independent real-world cohorts or can rediscover known clinicogenomic associations.

In the medium term, combining generative models with federated or privacy-preserving learning could enable multi-institutional multimodal FM training without centralizing sensitive data, offering a pragmatic path towards larger and more diverse datasets [136, 154, 226].

*D. Global Generalization versus Local Adaptation*

Building on the challenges outlined in Sections 6.1, 6.2 and the FM taxonomy in Section 5.4, this review considers large, multimodal cancer FMs trained across institutions and countries as a promising research direction, rather than a settled solution, for future oncology AI. However, real-world deployment will rarely be truly "universal." In practice, most hospitals operate within comparatively homogeneous environments: a limited set of scanner vendors and protocols in radiology, locally standardized staining and workflows in pathology, and institution-specific patient demographics and referral patterns. Domain shift between centers has been repeatedly shown to degrade model performance in imaging AI [227] even for relatively simple chest X-ray and CT tasks, so models validated at one site often underperform when deployed "as is" at another. Critically, this hierarchical view of global backbones and local adapters remains a working hypothesis. It will require formal testing in prospective, multi-site studies to determine whether such FM-based strategies consistently outperform simpler, locally optimized models, and under which deployment conditions, their additional complexity is justified.

A pragmatic view is therefore to treat cancer FMs as backbones rather than monolithic end-products. Pretraining should exploit as much cross-site diversity as governance allows us to learn broad, transferable representations, whereas deployment will typically require local specialization: lightweight fine-tuning or adapter modules on a hospital's own data, site-specific normalization and calibration layers, and, where direct data pooling is impossible, federated or multi-site fine-tuning protocols [136]. In this framing, the key design question is not "global or local?" but how best to combine global robustness (learned from heterogeneous, multi-institutional cohorts) with local optimization (to match specific equipment, workflows, and populations) for tasks such as image-based diagnosis, genomic interpretation, and prognostic modelling.

Practically, this points toward hierarchies of models: a shared, multimodal cancer FM pretrained across centers; institution- or region-level variants adapted via domain adaptation and site-specific adapters; and, in some cases, task-specific heads tuned to specific clinical endpoints. Such a tiered strategy acknowledges regulatory and operational realities while preserving the scientific value of globally pretrained backbones.

*E. Continual Learning*

A crucial next step for cancer FMs is to move beyond one-off training and periodic offline updates toward continual learning frameworks that can absorb new data, tasks, and clinical knowledge without catastrophic forgetting [228]. In practice, this means treating the FM as a living backbone that is incrementally updated as institutions acquire new imaging protocols [229], therapies, and patient populations, while preserving performance on prior cohorts. Techniques such as regularization-based approaches (e.g., elastic weight consolidation), replay or rehearsal of curated reference cohorts, and modular adapter layers offer complementary strategies to balance plasticity and stability in this setting. For oncology, continual learning is particularly important for modelling shifting standards of care (e.g., new immunotherapies), evolving resistance patterns, and changing screening practices, so that deployed FMs remain calibrated, clinically aligned, and empirically valid over time rather than becoming static "snapshots" of past practice.

## 7. Conclusions

To conclude, the integration of multimodal data with advanced AI holds an undeniable and revolutionary promise for oncology. Yet, the path from computational proof-of-concept to meaningful clinical impact is fraught with profound and deeply interconnected challenges. This review has systematically deconstructed these hurdles, revealing that they are not isolated issues to be solved in sequence, but a complex, interdependent system requiring a holistic and synergistic strategy. Our analysis reveals that the advancement of AI models is fundamentally reshaping the analysis of multi-modal and multimodal data. These state-of-the-art models excel at creating unified data representations, uncovering complex biological interactions, and delivering robust, scalable solutions for critical tasks like cancer subtyping, biomarker discovery, and survival prediction.

Clinically, these technological advancements are directly enhancing the translational pipeline, moving precision oncology from an aspirational goal to a clinical reality in a few years. The integrative methods reviewed here are yielding tangible improvements in early and non-invasive diagnostics, enabling more accurate prognostication, and personalizing therapeutic strategies by predicting patient response to specific treatments.

Despite this rapid progress, significant hurdles remain in the path to routine clinical deployment. The field must urgently address the need for standardized, multi-institutional validation protocols to ensure model robustness and generalizability, overcome challenges related to data harmonization, and enhance model interpretability to build clinical trust. Future efforts must be intensely focused on bridging the gap between computational innovation and real-world clinical utility. This will require fostering deep collaboration between data scientists and clinicians, promoting the development of accessible open-source tools, and establishing clear regulatory pathways to ensure that these transformative technologies can be safely and effectively integrated into patient care, ultimately realizing the promise of data-driven, personalized oncology.

**Author Contributions:**

A. M.: Conceptualization, Visualization, Writing – original draft, Writing – review and editing.

M. W.: Conceptualization, Writing – original draft, formal analysis, investigation

M. B. S.: Formal analysis, Visualization.

E. S.; R. B.; W. L: Formal analysis.

H. X.: Visualization.

J.C.; Z.L.; C.H.; L.S.S.; C.W.; N.V.; X.L.; L.B.; D.G.; J.H.; J.Z: Writing – review and editing

J. W.: Conceptualization, Writing - review and editing, Supervision.

**Funding:** This work was supported by generous philanthropic contributions to the MD Anderson Lung Cancer Moon Shot program as well as by the NIH/NCI under award number P30CA016672. This research was partially supported by NIH grants R01CA262425 and R01CA276178, as well as CPRIT RP240117 and NIH U24CA224285. This work was supported by the Tumor Measurement Initiative through the MD Anderson Strategic Initiative Development Program, Permanent Health Funds, and QIAC Partnership in Research Grant. Furthermore, this work was supported by generous philanthropic contributions from Andrea Mugnaini and Edward L. C. Smith. Finally, this work was supported by Rexanna's Foundation for Fighting Lung Cancer.

**Conflicts of Interest:** The authors declare no conflicts of interest.